\documentclass[a4paper,11pt]{article}
\pdfoutput=1
\usepackage{jheppub}
\usepackage[T1]{fontenc}
\usepackage{amsmath,amssymb}
\usepackage{epsfig}
\usepackage{graphicx}
\usepackage[usenames,dvipsnames]{color}
\usepackage{slashed}
\usepackage{float}
\usepackage{verbatim}

\newcommand{\be}{\begin{equation}}
\newcommand{\ee}{\end{equation}}

\newcommand{\om} {\omega}
\newcommand{\ob} {\bar{\omega}}
\newcommand{\rs} {\boldsymbol{1}}
\newcommand{\rw} {\boldsymbol{\omega}}
\newcommand{\rws} {\boldsymbol{\bar{\omega}}}
\newcommand{\rt} {\boldsymbol{3}}
\newcommand{\phil} {\phi_l}
\newcommand{\phis} {\phi_s}
\newcommand{\ks} {\kappa_s}
\newcommand{\ka} {\kappa_a}
\newcommand{\edit} {}

\begin{document}
\title{\boldmath Realization of the minimal extended seesaw mechanism and the $TM_2$ type neutrino mixing}

\author[a]{R. Krishnan,}
\author[b]{Ananya Mukherjee}
\author[b]{and Srubabati Goswami}

\affiliation[a]{Saha Institute of Nuclear Physics, 1/AF Bidhannagar, Kolkata 700064, India}
\affiliation[b]{Physical Research Laboratory, Ahmedabad- 380009, India}

\emailAdd{krishnan.rama@saha.ac.in}
\emailAdd{ananya@prl.res.in}
\emailAdd{sruba@prl.res.in}

\abstract{We construct a neutrino mass model based on the flavour symmetry group $A_4\times C_4 \times C_6 \times C_2$ which accommodates a light sterile neutrino in the minimal extended seesaw (MES) scheme. Besides the flavour symmetry, we introduce a $U(1)$ gauge symmetry in the sterile sector and also impose CP symmetry. The vacuum alignments of the scalar fields in the model spontaneously break these symmetries and lead to the construction of the fermion mass matrices. With the help of the MES formulas, we extract the light neutrino masses and the mixing observables. In the active neutrino sector, we obtain the $\text{TM}_2$ mixing pattern with non-zero reactor angle and broken $\mu$-$\tau$ reflection symmetry. We express all the active and the sterile oscillation observables in terms of only four real model parameters. Using this highly constrained scenario we predict $\sin^2 \theta_{23} =0.545^{+0.003}_{-0.004}$, $\sin \delta  =  -0.911^{+0.006}_{-0.005}$, $|U_{e4}|^2 = 0.029^{+0.009}_{-0.008}$, $|U_{\mu4}|^2 = 0.010^{+0.003}_{-0.003}$ and $|U_{\tau4}|^2 = 0.006^{+0.002}_{-0.002}$ which are consistent with the current data.}

% \pacs{14.60.Pq, 11.30.Qc}% PACS, the Physics and Astronomy
%                              % Classification Scheme.
% \keywords{Discrete flavour symmetry}%Use showkeys class option if keyword
%                               %display desired
\maketitle
\section{Introduction}
\label{sec1} 
Observations made in the neutrino oscillation experiments have confirmed that neutrinos have mass, albeit tiny. The Standard Model (SM) of particle physics can not accommodate the neutrino mass due to the absence of right-handed neutrinos, unlike the case for the charged leptons and the quarks. The inclusion of additional right-handed neutrino fields along with the seesaw mechanism \cite{Minkowski:1977sc,GellMann:1980vs,Yanagida:1979as,Mohapatra:1979ia} plays a vital role in modeling properties of massive neutrinos. The well known PMNS matrix encodes the mixing between the neutrino flavour eigenstates and their mass eigenstates. This matrix is parametrised in terms of three mixing angles and three CP phases (in a three flavoured paradigm),
\be\label{PMNS}
U_{\text{PMNS}}=\left(\begin{array}{ccc}
c_{12}c_{13}& s_{12}c_{13}& s_{13}e^{-i\delta}\\
-s_{12}c_{23}-c_{12}s_{23}s_{13}e^{i\delta}& c_{12}c_{23}-s_{12}s_{23}s_{13}e^{i\delta} & s_{23}c_{13} \\
s_{12}s_{23}-c_{12}c_{23}s_{13}e^{i\delta} & -c_{12}s_{23}-s_{12}c_{23}s_{13}e^{i\delta}& c_{23}c_{13}
\end{array}\right).\,U_{\text{Maj}},
\ee
where $c_{ij} = \cos{\theta_{ij}}, \; s_{ij} = \sin{\theta_{ij}}$. The diagonal
matrix, $U_{\text{Maj}}=\text{diag}(1, e^{i\alpha}, e^{i(\beta+\delta)})$, contains the 
Majorana CP phases $\alpha, \beta$ which become observable if the neutrinos behave as Majorana particles. 

Although the last two decades of neutrino oscillation experiments made tremendous progress in determining the three flavour mixing angles, efforts are underway to measure these parameters more precisely. We do not yet know whether the atmospheric mixing is maximal or not. If it is not, the octant of the atmospheric mixing angle, $\theta_{23}$, is to be determined. Measurement of the Dirac CP phase, $\delta$, will confirm CP violation in the leptonic sector and may explain the observed baryon asymmetry via leptogenesis. The nature of the neutrinos, i.e.~whether they are Dirac or Majorana, is still an open question which can not be settled with the help of the oscillation experiments. On the other hand, the observation of neutrino-less double-beta decays ($0\nu\beta\beta$) will establish the Majorana nature. Such decays are yet to be observed. The oscillation experiments have determined the mass-squared differences (solar: $\Delta m_{21}^2$ and atmospheric: $\Delta m_{31}^2$), but they are not sensitive to the absolute neutrino mass scale. Data from the Planck satellite provides an upper bound on the sum of neutrino masses, $\sum_i m_i \leq 0.16 $~eV~\cite{Aghanim:2018eyx}. Experimental searches are also being made to directly measure the electron neutrino mass using the kinematics of beta decays. Recently, the KATRIN collaboration has announced its first result on the effective electron antineutrino mass using the tritium beta decay, $^3\text{H} \rightarrow$ $^3\text{He} + e^- + \bar \nu_{\text{e}}$, and reported the upper bound for the effective antineutrino mass~\cite{Aker:2019qfn}, $m_{\bar \nu_e} < 1.1 $~eV at the $90\%$ confidence level (CL).

Although the three flavour paradigm of neutrino oscillation is well established, there are some experimental results that motivate us to go beyond this and postulate the existence of one or more sterile neutrinos. This possibility has gained considerable attention in recent years. In principle, the presence of a fourth neutrino can impressively explain several sets of experimental anomalies. The first indication came from the LSND experiment which showed evidence of oscillation with mass scale $\sim\text{eV}^2$ \cite{Athanassopoulos:1996jb} in $\bar{\nu}_\mu\text{-}\bar{\nu}_e$ channel. Later MiniBooNE experiment also confirmed it~\cite{Aguilar-Arevalo:2012fmn,Aguilar-Arevalo:2018gpe}. The \textit{Reactor Anomaly} involves a deficit of reactor antineutrinos detected in short-baseline (<500 m) experiments with recalculated neutrino fluxes~\cite{Mention:2011rk,Mueller:2011nm,Huber:2011wv}. The short-baseline neutrino oscillations can also explain the so-called \textit{Gallium Anomaly}  observed during the calibrations runs of the radiochemical experiments, GALLEX and SAGE. The ratio of the experimental flux to the theoretical estimate was found to be $0.86\pm 0.05$. The resolution of both the Reactor and the Gallium anomalies with the help of the active-sterile oscillations point towards a common region of the parameter space with the sterile neutrino having mass in the $\sim$~eV scale \cite{Dentler:2018sju,Abdurashitov:2005tb,PhysRevC.83.065504,Kostensalo:2019vmv}. 

The proposed sterile neutrino is an SM singlet which does not participate in the weak interactions, but they can mix with the active neutrinos enabling them to be probed in the oscillation experiments. The addition of a single sterile neutrino field leads to an oscillation parameter space consisting of a $4\times 4$ unitary mixing matrix along with three independent mass-squared-differences. Among them, the preferred scenario, often called the 3+1 scheme~\cite{Goswami:1995yq,Kopp:2011qd,Conrad:2012qt,Giunti:2011gz}, has three active neutrinos and one sterile neutrino in the sub-eV and eV scale respectively. The 2+2 scheme, in which two pairs of neutrino mass states differ by $\mathcal{O}(\text{eV})$, is not consistent with the solar and the atmospheric data \cite{Maltoni:2002ni}. The 1+3 scheme in which the three active neutrinos are in eV scale and the sterile neutrino is lighter than the active neutrinos is disfavored by cosmology. Therefore, in this paper, we assume the 3+1 scenario. The recently proposed Minimal Extended Seesaw (MES)~\cite{Barry:2011wb,Zhang:2011vh} has many appealing features. The active-sterile mixing obtained in MES is suppressed by the ratio of masses of the active and sterile sectors. With the active neutrino mass of the order of $\sim0.01$~eV and the sterile neutrino mass of the order of eV, this suppression is consistent with the active-sterile mixing as observed in LSND and MiniBooNe \footnote{The  data from solar and atmospheric neutrino oscillations as well as oscillations observed in accelerator experiments like T2K, MINOS, NOvA  and reactor experiments KamLAND and Daya-Bay, RENO, Double-Chooz etc. can be explained in terms of the three neutrino framework. Because dominant oscillations to sterile neutrinos is  disfavored as a solution to solar and atmospheric neutrino anomalies, the 2+2 picture is disfavored. In the 3+1 picture, the oscillations to  sterile neutrinos is a sub-leading effect to the dominant 3 flavour oscillations and the 3 generation global fit results are not altered.  The short baseline $eV^2$ oscillations can be explained using the One-mass-scale-Dominance approximation governed by $\Delta m^2_{41}$. However, there is a tension between observance of non-oscillation in   disappearance experiments and observation of oscillation in LSND and MiniBOONE which  makes the goodness of fit in the 3+1 picture worse.}

A large number of neutrino mass models based on discrete flavour symmetry groups have been proposed \cite{King:2015aea,Altarelli:2010gt,Smirnov:2011jv,King:2013eh} in the last decade. These models generate various mixing patterns such as the well known tribimaximal mixing (TBM)~\cite{Harrison:1999cf,Harrison:2002er,Harrison:2003aw,Altarelli:2005yx,Ma:2004zv,Ma:2005sha,Zee:2005ut}. Since the non-zero value of the reactor mixing angle ~\cite{Ahn:2012nd,Abe:2011fz,An:2016ses} has ruled out TBM, one of the popular ways to achieve realistic mixings is through either its modifications or extensions~\cite{King:2011zj, Antusch:2011ic, King:2011ab, Gupta:2011ct, Xing:2010pn, Harrison:2014jqa,Merle:2014eja}. Unlike the active-only mixing scenarios, realising the minimal extended type-I seesaw with the help of discrete groups is somewhat recent and limited ~\cite{Nath:2016mts,Das:2018qyt,Sarma:2018bgf,Dev:2012bd}. It is in this context that we propose a model to implement the MES and obtain oscillation observables consistent with the latest experiments. Our model produces an extension of the TBM called the $\text{TM}_2$~  \cite{Xing:2006xa,Albright:2008rp,Albright:2010ap,Ge:2011qn,Krishnan:2012me,Krishnan:2012sb,Vien:2016tmh,Krishnan:2018tja,Krishnan:2019ftw} in which the second column of the TBM is preserved. We use $A_4\times C_4$ as the flavour group for our model. We propose several scalar fields, often called the flavons, which couple with the charged-lepton fields as well as the various neutrino fields. The inherent properties of $A_4$ and $C_4$ as well as the residual symmetries of the vacuum alignments of the flavons, determine the structure and the symmetries of the mass matrices.   

The content of this paper is organised as follows. The features of the MES scheme are outlined in Section~\ref{sec2}. In Section~\ref{model}, we briefly explain the representation theory of the flavour group and move on to construct the Yukawa Lagrangian based on the proposed flavon content of the model. We also assign Vacuum Expectation Values (VEVs) for these flavons. {\edit We justify our assignments of VEVs with the help of symmetries in  Appendix~\ref{sec:appendixa}}. In Section~\ref{massmatrix}, the mass matrices are constructed in terms of the VEVs. We provide the formulae for various experimental observables as functions of the model parameters. In Section~\ref{pheno}, we compare these formulae with the experimental results and make predictions. {\edit We provide a representative set of model parameters in Appendix~\ref{sec:appendixb} and numerically extract the values of the observables so as to verify the validity of the various approximations used in the paper.} Finally, we conclude in Section~\ref{Conclusion}. 

\section{Minimal extended seesaw}\label{sec2}

In the Standard Model, the left-handed charged-lepton fields, $l_L = (e_L, \mu_L, \tau_L)^T$, and the neutrino fields, $\nu_L = (\nu_e, \nu_\mu, \nu_\tau)^T$, transform as the $SU(2)$ doublet, $L=(\nu_L, l_L)^T$. They couple with the right-handed charged-lepton fields, $l_R = (e_R, \mu_R, \tau_R)^T$, to form the charged-lepton mass term,
\be\label{eq:clmt}
\bar{L} y_l l_R H, 
\ee
where $y_l$ are the Yukawa couplings. In general, $y_l$ is a $3 \times 3$ complex matrix. The electroweak symmetry is spontaneously broken when the Higgs acquires the VEV,
\be\label{eq:higgsvev}
\langle H \rangle = (0, v)^T.
\ee
Subsequently, the mass term, Eq.~(\ref{eq:clmt}), becomes
\be \label{eq:clmm}
\bar{l}_L M_l l_R,
\ee
where $M_l = v y_l$ is the charged-lepton mass matrix. 

In the type-I seesaw framework, we add extra right-handed neutrino fields, $\nu_R$, to the SM. We may assume that three families of such fields exist, i.e.~$\nu_R = (\nu_{R1}, \nu_{R2}, \nu_{R3} )^T$. They couple with the left-handed fields, $L$, forming the Dirac neutrino mass term,
\be
\bar{L} y_\nu \nu_R \tilde{H},
\ee
where $\tilde{H} =i \sigma_2 {\edit H^*}$. As a result of the Spontaneous Symmetry Breaking (SSB), this term becomes
\be \label{eq:diracmm}
\bar{\nu}_L M_D \nu_R,
\ee
where $M_D = v y_\nu$ is the Dirac neutrino mass matrix. The right-handed neutrino fields can couple with themselves resulting in the Majorana mass term,
\be\label{eq:majoranamm}
\frac{1}{2}\bar{\nu}^c_R M_R \nu_R, 
\ee
where $M_R$ is the $3 \times 3$ Majorana neutrino mass matrix which is assumed to be at a very high scale in order to cause the seesaw suppression of the light neutrino masses. The canonical type-I seesaw can be extended to accommodate an eV-scale sterile neutrino at the cost of no fine-tuning of the Yukawa coupling. To implement this MES scheme we need to include an SM gauge singlet field, $\nu_s$, which couples with the heavy neutrino fields, $\nu_R$, leading to 
\be \label{eq:sterilemm}
{\edit\bar{\nu}^c_s M_s \nu_R},
\ee
where $M_s$ is a $1 \times 3$ mass matrix. We assume that the coupling of the sterile field ($\nu_s$) with itself as well as with the left-handed fields ($L$) is forbidden. 

Combining Eqs.~(\ref{eq:diracmm}, \ref{eq:majoranamm}, \ref{eq:sterilemm}), we obtain the Lagrangian containing the neutrino mass matrices relevant to the MES:
\be\label{eq:totalmm}
 \mathcal{L}_\nu = \bar \nu_L M_D \nu_R  + {\edit\bar{\nu}^c_s M_s \nu_R} + \frac{1}{2} \bar \nu_R^c M_R \nu_R+ h.c.
\ee
The Lagrangian, Eq.~(\ref{eq:totalmm}), leads to the following $7 \times 7$ neutrino mass matrix in the ($\nu_L, \nu_s^c, \nu_R^c$) basis:
\be\label{M7}
 M_\nu^{7 \times 7} = \left(\begin{array}{ccc}
       0 & 0 & M_D \\
       0 & 0 & {\edit M_s}\\
       M_D^T & {\edit M_s^T} & M_R
       \end{array}\right). 
\ee
Being analogous to the canonical type I seesaw, the MES scheme allows us to have the hierarchical mass spectrum assuming $M_R >> M_s > M_D$. The right-handed neutrinos are much heavier compared to the electroweak scale enabling them to be decoupled at the low scale. As a result, Eq.(\ref {M7}) can be block diagonalized to obtain the effective neutrino mass matrix in the ($\nu_L , \nu_s^c $) basis,
\be\label{eq:M4}
       M_\nu^{4 \times 4} = - \left(\begin{array}{cc}
       M_D M_R^{-1}M_D^T & M_D M_R^{-1} M_s^{T} \\
       M_s (M_R^{-1})^T M_D^T & M_s M_R^{-1}M_s^T 
       \end{array}\right).                 
\ee
This particular type of model is a minimal extension of the type I seesaw in the sense that
only an extra sterile field is added whose mass is also suppressed along with that of the three active neutrinos. Since $M_\nu^{7 \times 7}$ has rank 6 and subsequently $M_\nu^{4 \times 4}$ has rank three, the lightest neutrino state becomes massless\footnote{If we want to accommodate more than one sterile neutrino at the eV scale, we need to increase the number of heavy neutrinos as well. Otherwise, more than one active neutrino becomes massless which is ruled out experimentally.}.
   
Assuming $M_s > M_D$, we may apply a further seesaw approximation on Eq.(\ref {eq:M4}) to get the active neutrino mass matrix,
\be\label{eq:M3}
 M_\nu^{3\times 3} \simeq  M_D M_R^{-1}M_s^T (M_s M_R^{-1} M_s^T)^{-1} M_s M_R^{-1} M_D^T - M_D M_R^{-1} M_D^T.
\end{equation}
It is worth mentioning that the RHS of Eq. (\ref{eq:M3}) remains non-vanishing since $M_s$ is a row vector ${1\times3}$ rather than a square matrix. Under the approximation $M_s > M_D$, we also obtain the mass of the $4^\text{th}$ mass eigenstate\footnote{Since the active-sterile mixing is small, the $4^\text{th}$ mass eigenstate ($\nu_4$) more or less corresponds to the sterile state ($\nu_s$).},
\begin{equation}\label{eq:sterilemass}
 m_4 \simeq M_s M_R^{-1}M_s^T.
\ee

The charged-lepton mass matrix, $M_l$, Eq.~(\ref{eq:clmm}), is a $3\times 3$ complex matrix in general. Its diagonalisation leads to the charged-lepton masses,
\be
U_L M_l U_R^\dagger = \text{diag}(m_e, m_\mu, m_\tau),
\ee
where $U_L$ and $U_R$ are unitary matrices. The low energy effective $3\times 3$ neutrino mass matrix, $M_\nu^{3\times 3}$, Eq.~(\ref{eq:M3}), is complex symmetric. Its diagonalisation is given by
\be
U_\nu^\dagger M_\nu^{3\times 3} U_\nu^* = \text{diag}(m_1, m_2, m_3),
\ee
where $U_\nu$ is a unitary matrix and $m_1$, $m_2$ and $m_3$ are the light neutrino masses\footnote{In the MES framework, we have $m_1 = 0$.}. Using $U_L$ and $U_\nu$, we obtain the $4\times 4$ light neutrino mixing matrix, 
\be\label{eq:u44}
U \simeq \left(\begin{array}{cc}
       U_L(1-\frac{1}{2}R R^\dagger) U_\nu & U_L R \\
       -R ^\dagger U_\nu & 1-\frac{1}{2}R ^\dagger R 
       \end{array}\right) ,
\ee
where the three-component column vector $R$ is given by 
\be\label{Req}
  R = M_D M_R^{-1}M_s^T (M_s M_R^{-1}M_s^T)^{-1}.
\ee
$U$, Eq.~(\ref{eq:u44}), relates the neutrino mass eigenstates with the neutrino flavour eigenstates,
\be
U (\nu_1, \nu_2, \nu_3, \nu_4)^T = (\nu_e, \nu_\mu, \nu_\tau, \nu_s)^T,
\ee
in the basis where the charged-lepton mass matrix is diagonal. From Eq.~(\ref{eq:u44}), it is evident that the strength of the active-sterile mixing is governed by
\be\label{eq:sterileangles}
U_L R= (U_{e4}, U_{\mu 4}, U_{\tau 4})^T.
\ee
Note that $R$ is suppressed by the ratio $\mathcal{O}(M_D / M_s)$. The $3\times 3$ mixing matrix involving the three active neutrinos, $(\nu_e, \nu_\mu, \nu_\tau)$, and the three lightest mass eigenstates, $(\nu_1, \nu_2, \nu_3)$, is often called the PMNS mixing matrix, $U_\text{PMNS}$. In MES models with active-sterile mixing, $U_\text{PMNS}$ will not be unitary. It is given by the upper-left block of Eq.~(\ref{eq:u44}),
\be \label{eq:u33}
U_\text{PMNS} = U_L U_\nu-\frac{1}{2}U_LR R^\dagger U_\nu.
\ee
The deviation of $U_\text{PMNS}$ from unitarity, i.e.~$-\frac{1}{2}U_LR R^\dagger U_\nu$, is suppressed by $\mathcal{O}(M_D^2 / M_s^2)$.

%%%%%%%%%%%%%%%%%%%%%%%

\section{Flavour Structure of the Model}\label{model}
%\subsection{The flavour group: $A_4\times C_4$}

We construct the model in the framework of the discrete group $A_4 \times C_4$. $A_4$, which is the smallest group with a triplet irreducible representation, has been studied extensively in the literature \cite{Altarelli:2005yx,Ma:2001dn,Babu:2002dz,Shimizu:2011xg,Ishimori:2010au,Grimus:2011fk,King:2013eh,Altarelli:2010gt}. Here we briefly mention the essential features of this group in the context of model building. $A_4$ is the rotational symmetry group of the regular tetrahedron. It has the group presentation,
\be
\langle S,T ~|~ S^2 =T^3 = (ST)^3 = I \rangle.
\ee
$A_4$ has 12 elements which fall under four conjugacy classes. Its conjugacy classes and irreducible representations are listed in Table~\ref{tab:char}.

\begin{table}[h]
\begin{center}
\begin{tabular}{|c|c c c c|}
\hline
&$(1)$ & $(12)(34)$	& $(123)$ & $(132)$\\
\hline
$\rs$ & $1$ & $1$ & $1$ & $1$ \\
$\rw$ & $1$ & $1$ & $\om$ & $\ob$ \\
$\rws$ & $1$ & $1$ & $\ob$ & $\om$ \\
$\rt$ & $3$ & $-1$ & $0$ & $0$ \\
\hline
\end{tabular}
\caption{The character table of the $A_4$ group. $A_4$ denotes the even permutations of four objects. The conjugacy class (12)(34) represents two inversions carried out in two separate pairs of objects. (123) and (132) represent two inversions carried out in a set of three objects in the forward sense and the backward sense respectively. $\rs$ is the trivial representation. $\rw$ and $\rws$ are singlets transforming as $\om=e^{i\frac{2\pi}{3}}$ and $\ob=e^{-i\frac{2\pi}{3}}$ under $(123)$ and $(132)$. $\rt$ represents the three-dimensional rotational symmetries of a regular tetrahedron.}\label{tab:char}
\end{center}
\end{table}

For the triplet representation, $\rt$, we choose the following basis,
\be\label{eq:gens}
S =  \left(\begin{matrix} 1 & 0 & 0\\
0 & -1 & 0 \\
0 & 0 & -1
\end{matrix}\right), \quad T =  \left(\begin{matrix} 0 & 1 & 0\\
0 & 0 & 1 \\
1 & 0 & 0
\end{matrix}\right).
\ee
The representations $\rw$ and $\rws$ transform as $\om$ and $\ob$ respectively under the generator $T$ and trivially under the generator $S$. The tensor product of two triplets, $(x_1, x_2, x_3)$ and $(y_1, y_2, y_3)$, leads to 
\begin{align}
\rs& \equiv x_1 y_1 + x_2 y_2 + x_3 y_3\,, \label{eq:tp1}\\
\rw& \equiv x_1 y_1 + \ob x_2 y_2 + \om x_3 y_3\,, \label{eq:tpw}\\
\rws& \equiv x_1 y_1 + \om x_2 y_2 + \ob x_3 y_3\,, \label{eq:tpws}\\
\rt_s& \equiv (x_2 y_3+x_3 y_2, x_3 y_1+x_1 y_3, x_1 y_2+x_2 y_1)^T\,, \label{eq:tp3s}\\
\rt_a& \equiv (x_2 y_3-x_3 y_2, x_3 y_1-x_1 y_3, x_1 y_2-x_2 y_1)^T\,. \label{eq:tp3a}
\end{align}
The triplets $\rt_s$ and $\rt_a$, both of which transforming as $\rt$ under $A_4$, are constructed as the symmetric and the antisymmetric products respectively of $x$ and $y$.
%\subsection{Field assignments and the Lagrangian}
\begin{comment}
 \begin{table}[tbp]
\begin{center}
 \begin{tabular}{|c|c|c|c|c|c|c|c|c|c|c|c|c|c|c|}
\hline  
& $ L $ & $e_R$ & $\mu_R$ & $ \tau_R $ & $\nu_R$  & $ \nu_s $ & $\phil$ & $\eta$ & $\phi$ & $\phi_s$ & ${\edit \eta_\nu}$& $H$ & ${\edit H_s}$\\
\hline 
$A_4$ & $\rt$ & $\rs$ & $\rs$ & $\rs$ & $\rt$ & $\rs$  & $\rt$ & $\rs$ & $ \rt$ & $\rt$& ${\edit \rs}$ & $\rs$& ${\edit \rs}$\\
\hline     
$C_4$ & $1$ & $1$ & $1$ & $1$ & $-i$ & $i$  & $1$ & $i$ & $i$ & $1$& ${\edit -1}$ & $1$& ${\edit 1}$\\
\hline     
${\edit C_3}$ & ${\edit 1}$ & ${\edit1}$ & ${\edit \om}$ & ${\edit \ob}$ & ${\edit1}$ & ${\edit1}$  & ${\edit\om}$ & ${\edit1}$ & ${\edit1}$ & ${\edit1}$ & ${\edit1}$& ${\edit1}$& ${\edit1}$\\
\hline        
${\edit C_2\times U(1)_s}$ & ${\edit1}$ & ${\edit1}$ & ${\edit1}$ & ${\edit1}$ & ${\edit1}$ & ${\edit e^{iq \theta}}$  & ${\edit1}$ & ${\edit1}$ & ${\edit1}$ & ${\edit -1}$& ${\edit1}$ & ${\edit1}$& ${\edit -1\times e^{-i q\theta}}$\\
\hline   
 \end{tabular}
 \caption{The particle content and their charges under the flavour group of the model} \label{tab2}
\end{center}
\end{table} 
\end{comment}

 \begin{table}[tbp]
\begin{center}
 \begin{tabular}{|c|c|c|c|c|c|c|c|c|c|c|c|c|c|c|}
\hline  
& $ L $ & $e_R$ & $\mu_R$ & $ \tau_R $ & $\nu_R$  & $ \nu_s $ & $\phil$ & $\eta$ & $\phi$ & $\phi_s$ & $\eta_\nu$& $H$ & $H_s$\\
\hline 
$A_4$ & $\rt$ & $\rs$ & $\rs$ & $\rs$ & $\rt$ & $\rs$  & $\rt$ & $\rs$ & $ \rt$ & $\rt$& $\rs$ & $\rs$& $\rs$\\
\hline     
$C_4$ & $1$ & $1$ & $1$ & $1$ & $-i$ & $i$  & $1$ & $i$ & $i$ & $1$& $-1$ & $1$& $1$\\
\hline     
$C_6$ & $-\om$ & $-\om$ & $\ob$ & $1$ & $-\om$ & $-\om$  & $-\om$ & $1$ & $1$ & $\ob$ & $\om$& $1$& $\ob$\\
\hline        
$C_2\times U(1)_s$ & $1$ & $1$ & $1$ & $1$ & $1$ & $e^{iq \theta}$  & $1$ & $1$ & $1$ & $-1$& $1$ & $1$& $-1\times e^{-i q\theta}$\\
\hline   
 \end{tabular}
 \caption{The particle content and their charges under the flavour group of the model} \label{tab2}
\end{center}
\end{table} 
 
We extend the SM particle sector by the inclusion of three right-handed neutrinos, $\nu_{R}=(\nu_{R1}, \nu_{R2}, \nu_{R3})^T$, a sterile neutrino ($\nu_s$), several flavon multiplets, $\phil$, $\eta$, $\phi$, $\phis$, $\eta_\nu$ and a sterile sector Higgs, $H_s$. Along with the $A_4$ group, the model includes several Abelian discrete groups  ($C_4$, $C_6$ and $C_2$) acting variously on these fields. We also introduce a gauge group, $U(1)_s$, in the sterile sector. The field content of the model, along with the irreducible representations they belong to, are given in Table~\ref{tab2}. 
 
From these field assignments, we obtain the following Yukawa Lagrangian:
\begin{align}\label{eq:lagrangian}
\begin{split}
\mathcal{L}_Y =&Y_\tau \bar L\frac{ \phil}{\Lambda} \tau_R H +  Y_\mu \bar L\frac{ \phil^*}{\Lambda} \mu_R H+ \bar L\frac{Q}{\Lambda^2} e_R H\\
&\quad +Y_\eta\bar L \nu_R \frac{\eta}{\Lambda} \tilde{H} + Y_{\phi s} \left(\bar L \nu_R\right)^{\!T}_{\rt s} \frac{\phi}{\Lambda} \tilde{H}+ Y_{\phi a} \left(\bar L \nu_R\right)^{\!T}_{\rt a} \frac{\phi}{\Lambda} \tilde{H}\\
&\quad +  {\edit Y_s \frac{\phi_s^T}{\Lambda} \bar\nu_s^c \nu_R H_s}+ Y_{\nu}\eta_\nu \bar \nu_R^c \nu_R
\end{split}
\end{align}
where $Y_\tau$, $Y_\mu$, $Y_\eta$, $Y_{\phi s}$, $Y_{\phi a}$, $Y_s$, and $Y_\nu$ are the Yukawa-like dimensionless coupling constants, $\Lambda$ is the cut-off scale of the theory. $()_{\rt s}$ and $()_{\rt a}$ represent the symmetric and the antisymmetric tensor products given in Eq.~(\ref{eq:tp3s}) and Eq.~(\ref{eq:tp3a}) respectively. $Q$ represents all the quadratic flavon terms forming a triplet under $A_4$ and an invariant under the rest of the flavour group. It consists of $(\phil^*\phil)_{\rt s}$, $(\phil^*\phil)_{\rt a}$, $(\phis^*\phis)_{\rt s}$, $(\phis^*\phis)_{\rt a}$, {\edit $(\phi^*\phi)_{\rt s}$, $(\phi^*\phi)_{\rt a}$, $\eta^*\phi$, $\eta\phi^*$} along with the corresponding Yukawa-like coupling constants. Besides the flavour symmetries, we also impose the CP symmetry at high energy scales where the Higgses and the flavons have not acquired their VEVs, i.e. all the Yukawa-like couplings in the Lagrangian are real. We also assume that these couplings are of the order of one.
 
In the above Lagrangian, the first three terms are responsible for the charged-lepton mass generation. Since $\phil$, $\phil^*$ and $Q$ transform as $-\om$, $-\ob$ and $1$ under $C_6$, they couple with $\bar L \tau_R$, $\bar L \mu_R$ and $\bar L e_R$ (which transform as $-\ob$, $-\om$ and $1$) respectively. The rest of the terms involve the right-handed neutrino triplet $\nu_R$ and they contribute to the neutrino mass generation. The terms in the second line of Eq.~(\ref{eq:lagrangian}) contain $\bar L \nu_R$ which transforms as $-i$ under $C_4$ and remains invariant $C_6$. The flavons $\phi$ and $\eta$, which transform as $i$ under $C_4$ and remain invariant $C_6$, couple with $\bar L \nu_R$. These terms result in the Dirac mass matrix for the neutrinos. Under $A_4\times C_4\times C_6 \times C_2 \times U(1)_s$, the term $\bar \nu_s^c \nu_R$ transforms as $\rt\times 1\times \ob \times 1 \times e^{i q\theta}$. This term couples with $\phis$ and $H_s$ which transform as $\rt\times 1 \times \ob \times -1 \times 1$ and $\rs \times 1\times \ob \times -1 \times e^{-i q\theta}$ respectively and forms the sterile mass term.  Finally, we have the Majorana mass term consisting of $\bar \nu_R^c \nu_R$ coupled with the flavon $\eta_\nu$. We may also construct the neutrino mass terms $(\bar L \tilde{H}) (\tilde{H}^T L^c) \frac{Q_1}{\Lambda^3}$, $\bar L\nu_s\frac{Q_2}{\Lambda^3}\tilde{H}H_s$ and $\bar \nu_s^c \nu_s \frac{Q_3}{\Lambda^3} H_s^2$ where $Q_1$, $Q_2$ and $Q_3$ represent the quadratic flavon terms transforming as $1 \times \ob \times 1$, $-i \times \om \times -1$ and $-1 \times 1 \times 1$ respectively under $C_4 \times C_6 \times C_2$. Since these mass terms are heavily suppressed, we have not included them in the Lagrangian. In fact, a part of the reason for assigning the various Abelian charges to the flavons, Table~\ref{tab2}, is to ensure that this suppression occurs so that our model leads to the standard MES framework where these terms are assumed to vanish (the block of zeros in Eq.~(\ref{M7})).

Like the SM Higgs, the other scalar fields in the model also acquire VEVs through SSB. We assign them the following values:
\begin{align}
\langle \phil \rangle &= v_l(1,\ob,\om)^T,\label{eq:vevphil}\\
\langle \eta \rangle &= v_\eta,\label{eq:veveta}\\
\langle \phi \rangle &= v_\phi (0,-i,0)^T,\label{eq:vevphi}\\
\langle \phis \rangle &= v_s (1,0,1)^T,\label{eq:vevphis}\\
\langle \eta_\nu \rangle & = v_\nu,\label{eq:vevphinu}\\
\langle H_s \rangle &= v'.\label{eq:vevhs}
\end{align}
The VEVs of the various flavons in the model break the discrete flavour group, $A_4\times C_4 \times C_6 \times C_2$ in specific ways. In Appendix A, we study the residual symmetries of these VEVs and describe how their alignments can be uniquely defined.

The VEV of the sterile Higgs ($H_s$) breaks the $U(1)_s$ gauge group and leads to a massive gauge boson. This particle can mediate the so-called secret interactions of the sterile neutrinos proposed in the cosmological context~\cite{Dasgupta:2013zpn,Chu:2018gxk,Mazumdar:2019tbm}. The VEV of the sterile Higgs is assumed to be an order of magnitude higher than the VEV of the SM Higgs, i.e.~$v=176$~GeV, $v'\approx2000$~GeV. We also assume that the flavon VEVs are at a very high energy scale~$v_x\approx10^{10}$~GeV where $v_x$ denotes $v_l$, $v_\eta$, $v_\phi$, $v_s$ and $v_\nu$ and that the cut-off scale $\Lambda\approx10^{13}$~GeV. Under these assumptions, we may calculate the scales of our mass terms (after SSB); $\bar l_L \tau_R, \, \bar l_L \mu_R:$~$v \frac{v_l}{\Lambda} \approx10^{-1}$, $\bar l_L e$:~$v \frac{\mathcal{O}(v_x^2)}{\Lambda^2} \approx 10^{-4}$, $\bar \nu_L \nu_R$:~$v \frac{\mathcal{O}(v_x)}{\Lambda} \approx 10^{-1}$, $\bar \nu_s^c \nu_R$:~$v' \frac{v_s}{\Lambda} \approx 1$, $\bar \nu_R^c \nu_R$:~$v_\nu\approx 10^{10}$, $\bar \nu_L \nu_L^c$:~$v^2 \frac{\mathcal{O}(v_x^2)}{\Lambda^3}\approx 10^{-15}$, $\bar \nu_L \nu_s$:~$v v' \frac{\mathcal{O}(v_x^2)}{\Lambda^3}\approx 10^{-14}$ and  $\bar \nu_s^c \nu_s$:~$v'^2 \frac{\mathcal{O}(v_x^2)}{\Lambda^3}\approx 10^{-13}$ where all the units are in GeV.

\section{Mass Matrices and Observables}\label{massmatrix}

Substituting the Higgs VEV and the flavon VEVs in the Lagrangian for the charged-lepton sector (first line of Eq.~(\ref{eq:lagrangian})), we obtain the charged-lepton mass matrix, Eq.~(\ref{eq:clmm}),
 \be\label{CL}
       M_l = v \frac{\mathcal{O}(v_x^2)}{\Lambda^2}\left(\begin{array}{ccc}
       \mathcal{O}(1) & 0 & 0 \\
       \mathcal{O}(1) & 0 & 0  \\ 
       \mathcal{O}(1) & 0  & 0
       \end{array}\right)+v \frac{v_l}{\Lambda}\left(\begin{array}{ccc}
       0 & Y_\mu & Y_\tau \\
       0 & \om Y_\mu & \ob Y_\tau  \\ 
       0 & \ob Y_\mu  & \om Y_\tau
       \end{array}\right).          
\ee
This mass matrix is diagonalised using the transformation,
\begin{equation}\label{eq:cldiagonalisation}
U_L \, M_l \,U_R^\dagger=\text{diag}(m_e, m_\mu, m_\tau),
\end{equation}
where 
\be\label{eq:uomega}
      U_L \simeq \frac{1}{\sqrt{3}}\left(\begin{array}{ccc}
       1 & 1 & 1 \\
       1 & \ob & \om \\ 
       1 & \om & \ob
       \end{array}\right),
\ee
$U_R$ is an unobservable unitary matrix and 
\be \label{eq:clmasses}
m_e = v \frac{\mathcal{O}(v_x^2)}{\Lambda^2}, \quad m_\mu \simeq \sqrt{3}Y_\mu v \frac{v_l}{\Lambda}, \quad m_\tau \simeq \sqrt{3}Y_\tau v \frac{v_l}{\Lambda}
\ee
are the masses of the charged leptons. Here, the electron mass is suppressed by a factor of $\frac{v_x}{\Lambda}\approx10^{-3}$ compared to the muon and the tau masses. This is similar to the Froggatt-Nielsen mechanism which was proposed to explain the hierarchy of the fermion masses. It can be shown that the error in the expression of $U_L$ given in Eq.~(\ref{eq:uomega}) is of the order of $\frac{v_x^2}{\Lambda^2}$.  The relative errors in the expressions of $m_\mu$ and $m_\tau$ given in Eqs.~(\ref{eq:clmasses}) are also of the same order. These errors are very small and hence can safely be ignored. For a numerical verification, please refer to Appendix~B.

The terms in the second line of the Lagrangian, Eq.~(\ref{eq:lagrangian}), generate the Dirac mass matrix for the neutrinos. Substituting Higgs VEV and the VEVs of the flavons $\eta$ and $\phi$, Eqs.~(\ref{eq:veveta}, \ref{eq:vevphi}), in these terms, we obtain the Dirac neutrino mass matrix, Eq.~(\ref{eq:diracmm}),
\be\label{eq:dirac}
       M_{D} = v \frac{1}{\Lambda}\left(\begin{array}{ccc}
       v_\eta Y_\eta & 0 & -i v_\phi (Y_{\phi s}-Y_{\phi a})\\
       0 & v_\eta Y_\eta & 0  \\ 
       -i v_\phi (Y_{\phi s}+Y_{\phi a}) & 0  & v_\eta Y_\eta
       \end{array}\right).                
\ee

Substituting the VEV of $\phis$, Eq.~(\ref{eq:vevphis}), in the mass term for the sterile neutrino, $Y_s \bar \nu_R^c \nu_s \frac{\phi_s}{\Lambda}H_s$, we obtain the mass matrix representing the couplings between $\nu_s$ and $\nu_R$, Eq.~(\ref{eq:sterilemm}),
\be\label{eq:sterile}
M_s = v' \frac{v_s}{\Lambda} Y_s  {\edit (1, 0, 1)}.             
\ee
Finally, from the term $Y_{\nu}\eta_\nu \bar \nu_R^c \nu_R$, we obtain the mass matrix for the heavy right-handed neutrinos, Eq.~(\ref{eq:majoranamm}),
\be\label{eq:majorana}
M_R = Y_\nu v_\nu I,
\ee
where $I$ is the $3\times3$ identity matrix.

We implement the MES scheme, Eq. (\ref {eq:M4}), using the neutrino mass matrices, $M_D$, $M_s$, $M_R$, Eqs.~(\ref{eq:dirac}, \ref{eq:sterile}, \ref{eq:majorana}), and obtain the following effective neutrino mass matrix:
    \be \label{eq:m44model}
  M_\nu ^{4 \times 4} =\left( 
\begin{array}{c@{}c}
m\left(\begin{array}{ccc}
                       1-(\ks-\ka)^2 &0 & -2i\ks\\ 
                       0 & 1 & 0\\
                       -2i\ks & 0 & 1-(\ks+\ka)^2\\
                      \end{array}\right) & \frac{\sqrt{m m_s}}{\sqrt{2}}\left(\begin{array}{c}
                     1-i(\ks-\ka)\\ 
                     0\\
                     1-i(\ks+\ka)\\
                      \end{array}\right)\\
\frac{\sqrt{m m_s}}{\sqrt{2}} \left(\begin{array}{ccc}
                       1-i(\ks-\ka) & 0 & 1-i(\ks+\ka)\\ 
                      \end{array}\right) & \quad\quad\quad m_s \\
\end{array}\right),
\ee
where
\be
m=\frac{v^2 v_\eta^2 Y_\eta^2}{Y_{\nu}v_\nu\Lambda^2}, \quad m_s =  \frac{2 v'^2v_s^2 Y_s^2}{Y_{\nu}v_\nu\Lambda^2}, \quad \ks = \frac{v_\phi Y_{\phi s}}{v_\eta Y_\eta}, \quad \ka = \frac{v_\phi Y_{\phi a}}{v_\eta Y_\eta}.
\ee  
Here, the mass $m$ is suppressed by the very high value of $v_\nu$ ($\approx 10^{10}$~GeV) and also by the ratio $\frac{\Lambda^2}{v_\eta^2}$ ($\approx 10^6$). Hence, we obtain $m$ at around $0.01$~eV. The ratio $\frac{m}{m_s}$ relates the scale of the active sector of the neutrinos to that of the sterile sector and it is given by $\frac{m}{m_s} =\frac{v^2 v_\eta^2Y_\eta^2}{2v'^2v_s^2 Y_s^2}$. Since $v'$ is assumed to be an order of magnitude higher than $v$, the ratio $\frac{m}{m_s}$ becomes small ($\approx 0.01$). As a result, we can use Eq.~\ref{eq:M3} to obtain the effective $3\times 3$ neutrino mass matrix,                                  
 \begin{equation}\label{eq:mes}
M _\nu^{3\times3} = \frac{m}{2}\left(\begin{array}{ccc}
    (\ks-\ka-i)^2 & 0 & \ka^2 - (\ks-i)^2\\
 0 & -2 & 0 \\
 \ka^2 - (\ks-i)^2 & 0 & (\ks+\ka-i)^2 
   \end{array}\right).
\end{equation}
Using the unitary matrix,
\be \label{eq:unu}
U_\nu = \frac{1}{\sqrt{2} \kappa}\left(\begin{array}{ccc}
       i+\ks+\ka & 0 &  -i+\ks-\ka\\
       0 & i\sqrt{2}\kappa  & 0 \\ 
       i+\ks-\ka & 0 & i-\ks-\ka 
       \end{array}\right)\quad \text{with} \quad \kappa=\sqrt{(1+\ks^2+\ka^2)},
\ee
we diagonalise $M _\nu^{3\times3}$, Eq.~(\ref{eq:mes}),
\be
U_\nu^\dagger M_\nu^{3\times3} U_\nu^* = m\,\,\text{diag}\left(0, 1, 1+\ks^2+\ka^2  \right),
\ee
to obtain the light neutrino masses,
\be \label{eq:masses}
m_1 = 0, \quad m_2 = m, \quad m_3 = m (1+\ks^2+\ka^2).
\ee

Using the expressions of $U_L$ and $U_\nu$, Eqs.~(\ref{eq:uomega}, \ref{eq:unu}), we obtain the PMNS mixing matrix ($U_\text{PMNS} \simeq U_L U_\nu$\footnote{ $U_\text{PMNS}$ obtained in this approximation is unitary. For a numerical analysis without this approximation which leads to non-unitarity, please refer to Appendix~B.}, Eq.~(\ref{eq:u33})) in terms of the parameters $\ks$ and $\ka$,
\be\label{eq:modelpmns}
U_\text{PMNS} \simeq \frac{1}{\sqrt{6}\kappa}\left(\begin{array}{ccc}
     2(i+\ks) & i\sqrt{2}\kappa&  -2 \ka \\
       (i+\ks)(1+\om)+\ka(1-\om) & i\sqrt{2}\kappa \ob & (-i+\ks)(1-\om)-\ka(1+\om) \\ 
       (i+\ks)(1+\ob)+\ka(1-\ob) & i\sqrt{2}\kappa \om & (-i+\ks)(1-\ob)-\ka(1+\ob)
       \end{array}\right).
\ee
The absolute values of the elements of the middle column of this mixing matrix are equal to $\frac{1}{\sqrt{3}}$, i.e.~the mixing has the $\text{TM}_2$ form. Note that the eigenvalue $m$ should correspond to the second neutrino eigenstate because this eigenstate should remain unmixed with the others in order to obtain the $\text{TM}_2$ mixing, Eq.~(\ref{eq:modelpmns}). Therefore, the mass ordering should be either $(0, m, m(1+{\edit \kappa_s^2}+{\edit \kappa_a^2}))$ or $(m(1+{\edit \kappa_s^2}+{\edit{\kappa_a^2}}), m, 0)$. The second case is inconsistent with the experimental observation of $m_1<m_2$ given that $\ks$ and $\ka$ are real parameters. Hence, the model predicts the normal ordering (the first case) of the light neutrino masses.

From Eq.~(\ref{eq:modelpmns}), we extract the three mixing angles in the active sector,
\begin{align}
\sin^2 \theta_{13}&=\frac{2 \ka^2}{3(1+\ks^2+\ka^2)},\label{eq:t13}\\
\sin^2 \theta_{12}&=\frac{1+\ks^2+\ka^2}{3+3 \ks^2+\ka^2},\label{eq:t12}\\
\sin^2 \theta_{23}&=\frac{3+3\ks^2+ 2\sqrt{3} \ka + \ka^2}{2(3+3 \ks^2+\ka^2)}.\label{eq:t23}
\end{align}
Eq.~(\ref{eq:t13}) and Eq.~(\ref{eq:t12}) are consistent with the $\text{TM}_2$ constraint, $\sin^2 \theta_{12} \cos^2 \theta_{13}=\frac{1}{3}$. Using Eqs.~(\ref{eq:masses}, \ref{eq:t13}-\ref{eq:t23}), we obtain another constraint among the observables,
\be\label{eq:atmconstraint}
\sin^2 \theta_{23}=\frac{1}{2}+\frac{3}{\sqrt{2}} \sin \theta_{13} \sin^2 \theta_{12} \left(\frac{\Delta m^2_{21}}{\Delta m^2_{31}}\right)^{\frac{1}{4}},
\ee
which shows the deviation from the maximal atmospheric mixing, i.e.~$\sin^2 \theta_{23}=\frac{1}{2}.$
We also calculate the Jalskog's CP-violation parameter \cite{Jarlskog:1985ht} in the active sector, 
\be\label{eq:jkska}
J=\text{Im}(U_{e2} U_{\mu 3} U^*_{e3} U^*_{\mu 2})=-\frac{\ks \ka}{3\sqrt{3}(1+\ks^2+\ka^2)}.
\ee
Eliminating $\kappa_s$ and $\kappa_a$ from Eq.~(\ref{eq:jkska}) using Eqs.~(\ref{eq:masses}, \ref{eq:t13}), we can express $J$ in terms of the reactor angle and the light neutrino masses,
\be
J=-\frac{1}{3\sqrt{2}} \sin \theta_{13}\sqrt{1-\frac{3}{2}\sin^2 \theta_{13}-\frac{\sqrt{\Delta m^2_{21}}}{\sqrt{\Delta m^2_{31}}}}.
\ee
Given the three mixing angles and J in terms of the model parameters, we can obtain $\sin\delta$
using the following expression:
\be\label{eq:sindelta}
\sin \delta = J/(\sin\theta_{13}\sin\theta_{12}\sin\theta_{23}\cos^2\theta_{13}\cos \theta_{12}\cos \theta_{23}).
\ee
Note that the $4\times 4$ mixing matrix is parametrised using six mixing angles ($\theta_{13}$, $\theta_{12}$, $\theta_{23}$, $\theta_{14}$, $\theta_{24}$, $\theta_{34}$) and three Dirac CP phases ($\delta_{13}$, $\delta_{14}$, $\delta_{24}$) with the help of the parametrization mentioned in Ref.~\cite{Barry:2011wb}. However, we used the parametrisation for the $3\times3$ mixing matrix, Eq.~(\ref{PMNS}), to extract the mixing angles ($\theta_{13}$, $\theta_{12}$, $\theta_{23}$) and the CP phase ($\delta=\delta_{13}$) given in Eqs.~(\ref{eq:t13}, \ref{eq:t12}, \ref{eq:t23}, \ref{eq:sindelta}). This approximation is valid since the active-sterile mixing is quite small.

Comparing Eqs.~(\ref{eq:M4}, \ref{eq:sterilemass}) with Eq.~(\ref{eq:m44model}), it is clear that the model parameter $m_s$ corresponds to the mass of the $4^\text{th}$ mass eigenstate,
\be\label{eq:m4ms}
m_4 = m_s.
\ee
 Using Eq.~(\ref{Req}), we obtain the three-component column vector $R$,
\be\label{eq:r}
 R = \sqrt{\frac{m}{2 m_s}}\left(\begin{array}{c}
       1-i(\ks-\ka)  \\
       0\\ 
       1-i (\ks + \ka)
       \end{array}\right).
\ee
Substituting Eqs.~(\ref{eq:uomega}, \ref{eq:r}) in Eq.~(\ref{eq:sterileangles}) we obtain
\begin{align}
&U_{e4} = \frac{\sqrt{2m}}{\sqrt{3m_s}}(1-i\ks),\label{eq:ue4}\\
&U_{\mu4} =  -\frac{\ob\sqrt{m}}{\sqrt{6m_s}}(1-i\ks +\sqrt{3} \ka),\label{eq:umu4}\\
&U_{\tau4} =  -\frac{\om\sqrt{m}}{\sqrt{6m_s}}(1-i\ks -\sqrt{3} \ka).\label{eq:utau4}
\end{align}
Using Eqs.~(\ref{eq:ue4}, \ref{eq:umu4}, \ref{eq:utau4}), we can write the three active-sterile mixing angles in terms of the model parameters,
\begin{gather}
 \text{sin}^2\theta_{14}= \frac{2}{3}\frac{m}{m_s}(1+\ks^2),\\
 \text{sin}^2\theta_{24}= \frac{1}{6}\frac{m}{m_s}\frac{(1+\ks^2+2\sqrt{3}\ka+3\ka^2)}{1- \frac{2}{3}\frac{m}{m_s}(1+\ks^2)},\\
 \text{sin}^2\theta_{34}= \frac{1}{6}\frac{m}{m_s}\frac{(1+\ks^2-2\sqrt{3}\ka+3\ka^2)}{1-\frac{2}{3}\frac{m}{m_s}(1+\ks^2)-\frac{1}{6}\frac{m}{m_s}(1+\ks^2+2\sqrt{3}\ka+3\ka^2)}.
\end{gather}

For the extraction of $\delta_{14}$ and $\delta_{24}$, we need to calculate the Jarlskog-like rephasing invariants from the active-sterile sector. In this context, we refer the readers to a recent work \cite{Reyimuaji:2019wbn}, in which nine independent rephasing invariants in terms of the six mixing angles and the three Dirac phases have been evaluated in the context of the $4\times 4$ mixing matrix. With the help of these invariants, we may extract $\delta_{14}$ and $\delta_{24}$.

The effective neutrino mass applicable to the neutrinoless double-beta decay~\cite{Bamert:1994qh,Benes:2005hn} is given by
\be
m_{\beta\beta}=\left|m_1 U_{e1}^2+m_2 U_{e2}^2+m_3 U_{e3}^2+m_s U_{e4}^2\right|.
\ee
Substituting the values of the neutrinos masses, Eq.~(\ref{eq:masses}), the elements of the first row of the mixing matrix, Eq.~(\ref{eq:modelpmns}), and the expression for $U_{e4}$, Eq.~(\ref{eq:ue4}), in the above equation, we obtain
\be\label{eq:mbb}
m_{\beta\beta} = \left|\frac{m}{3}(1-2\ks^2+2\ka^2-4i\ks)\right|=\frac{m}{3} \sqrt{(1-2\ks^2+2\ka^2)^2+16\ks^2}.
\ee

\section{Phenomenology and Predictions}\label{pheno}

\begin{table}[bp]
\begin{center}
\begin{tabular}{|c|c|}
\hline
& $3\sigma$ range\\
\hline
$\sin^2 \theta_{13}$ & $0.02044 \rightarrow 0.02437$ \\
$\Delta m^2_{21}$ & $6.79\times10^{-5}~\text{eV}^2 \rightarrow 8.01\times10^{-5}~\text{eV}^2$\\
$\Delta m^2_{31}$ & $2.431\times10^{-3}~\text{eV}^2 \rightarrow 2.622\times10^{-3}~\text{eV}^2$\\
$\Delta m^2_{41}$ & $0.87~\text{eV}^2 \rightarrow 2.04~\text{eV}^2$\\
\hline
\end{tabular}
\caption{The mixing observables which are used to evaluate the model parameters $\ks$, $\ka$, $m$ and $m_s$. The $3\sigma$ ranges of $\sin^2 \theta_{13}$, $\Delta m^2_{21}$ and $\Delta m^2_{31}$ are taken from Ref.~\cite{article} and that of $\Delta m^2_{41}$ is taken from Ref.~\cite{}.}
\label{tab:input}
\end{center}
\end{table}

Our model allows only four degrees of freedom in the neutrino Yukawa sector, denoted by the free parameters $\ks$, $\ka$, $m$, $m_s$. There are eleven independent experimentally measured quantities, $\sin^2 \theta_{12}$, $\sin^2 \theta_{23}$, $\sin^2 \theta_{13}$, $\sin \delta$, $\Delta m^2_{21}$, $\Delta m^2_{31}$, $m_{\beta\beta}$, $\Delta m^2_{41}$, $|U_{e4}|^2$, $|U_{\mu4}|^2$, $|U_{\tau4}|^2$ all of which can be expressed in terms of the above mentioned four model parameters. So, it is clear that the model is extremely constrained. In this section, we calculate the model parameters using the experimental data and also make predictions. To calculate the allowed ranges of the model parameters, $\ks$, $\ka$, $m$ and $m_s$, we utilize the observables $\sin^2 \theta_{13}$, $\Delta m^2_{21}$, $\Delta m^2_{31}$ and $\Delta m^2_{41}$ whose experimental values are given in Table~\ref{tab:input}. These values are obtained from the global fit data published by the nufit group \cite{article} and the active-sterile mixing data from Ref.~\cite{Gariazzo:2015rra}. 

Using the expression of the reactor mixing angle given in Eq.~(\ref{eq:t13}) and its range given in Table~\ref{tab:input}, we obtain
\be\label{eq:t13range}
0.02044 \leq \frac{2 \ka^2}{3(1+\ks^2+\ka^2)}  \leq 0.02437.
\ee

Using Eq.~(\ref{eq:masses}), we calculate the ratio of the mass-squared differences of the active neutrinos,
\be\label{eq:massratio}
\frac{\Delta m^2_{31}}{\Delta m^2_{21}} = (1+\ks^2+\ka^2)^2.
\ee
Given the ranges of $\Delta m^2_{21}$ and $\Delta m^2_{31}$  (Table~\ref{tab:input}), Eq.~(\ref{eq:massratio}) leads to
\be\label{eq:massratiorange}
30.3 \leq (1+\ks^2+\ka^2)^2  \leq 38.6.
\ee

\begin{figure}[h]
\begin{center}
\includegraphics[scale=0.70]{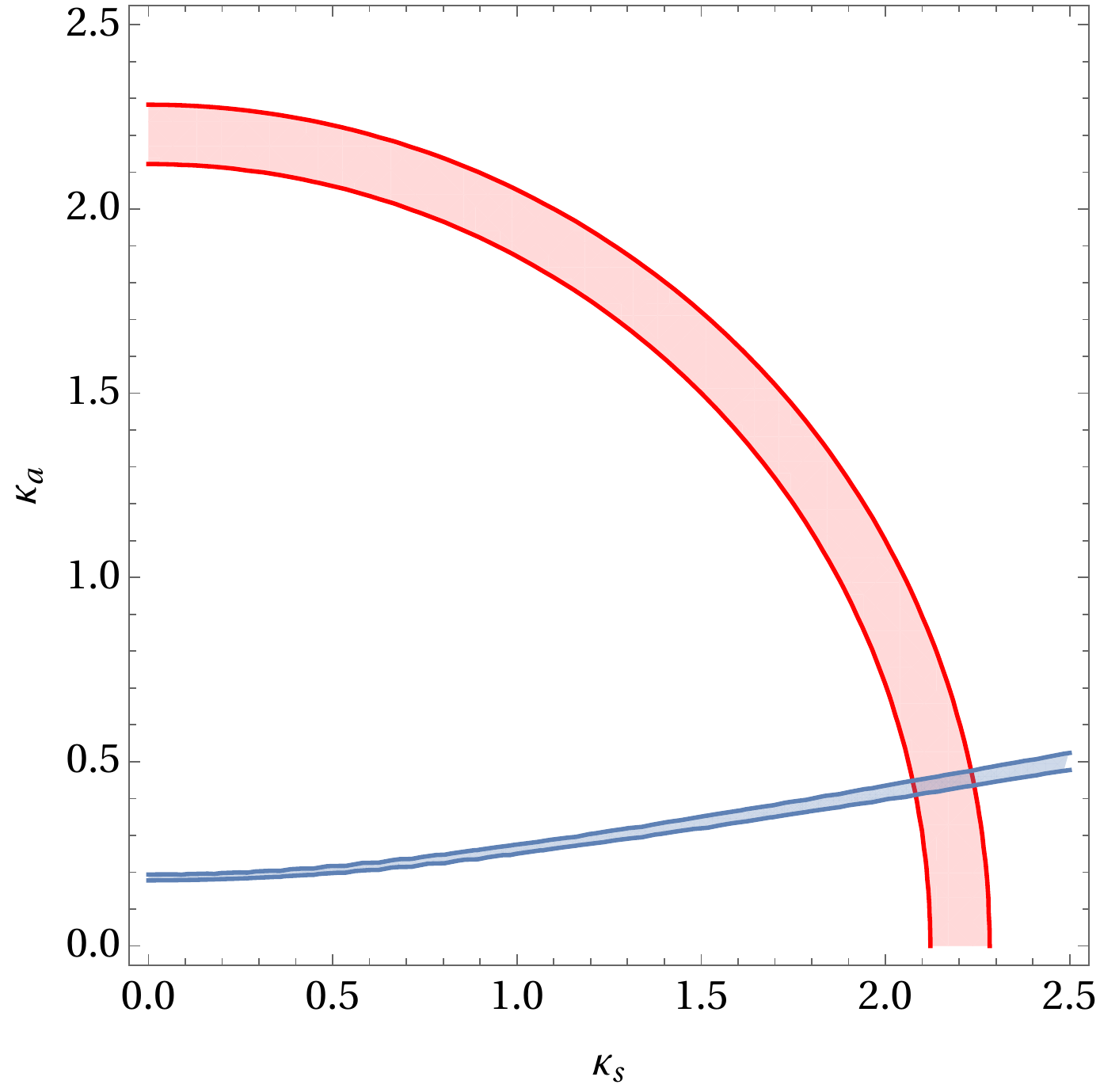}
\caption[]{The parameters $\ks$ and $\ka$ constrained using the reactor mixing angle and the ratio of the mass-squared differences of the active neutrinos.}
\label{fig:kska}
\end{center}
\end{figure}

We use Eqs.~(\ref{eq:t13range}, \ref{eq:massratiorange}) to constrain the parameters $\ks$ and $\ka$. In this analysis, we chose $\sin^2 \theta_{13}$ and $\frac{\Delta m^2_{31}}{\Delta m^2_{21}} $ because these are the most precisely measured observables that can be used for constraining $\ks$ and $\ka$. The results are shown in Figure~\ref{fig:kska} where the blue and the red regions represent the constraints Eq.~(\ref{eq:t13range}) and Eq.~(\ref{eq:massratiorange}) respectively. The allowed range of $\ks$ and $\ka$ is given by the intersection of the red and the blue regions. Note that these parameters can be directly expressed in terms of the observables:
\be\label{eq:ks2ka2}
{\edit \kappa_s^2}=\left(1-\frac{3}{2}\sin^2 \theta_{13}\right)\frac{\sqrt{\Delta m^2_{31}}}{\sqrt{\Delta m^2_{21}}}-1, \quad \quad
{\edit \kappa_a^2}=\frac{3}{2}\sin^2\theta_{13}\frac{\sqrt{\Delta m^2_{31}}}{\sqrt{\Delta m^2_{21}}}.
\ee
The best fit values, i.e.~$\sin^2 \theta_{13}=0.02237$, $\Delta m^2_{21}=7.39\times 10^{-5}~\text{eV}^2$ and $\Delta m^2_{31}=2.528\times 10^{-3}~\text{eV}^2$, leads to ${\edit \kappa_s}=2.16$ and ${\edit \kappa_a}=0.44$ consistent with Figure~\ref{fig:kska}.

\begin{figure}[]
\begin{center}
\includegraphics[scale=0.2]{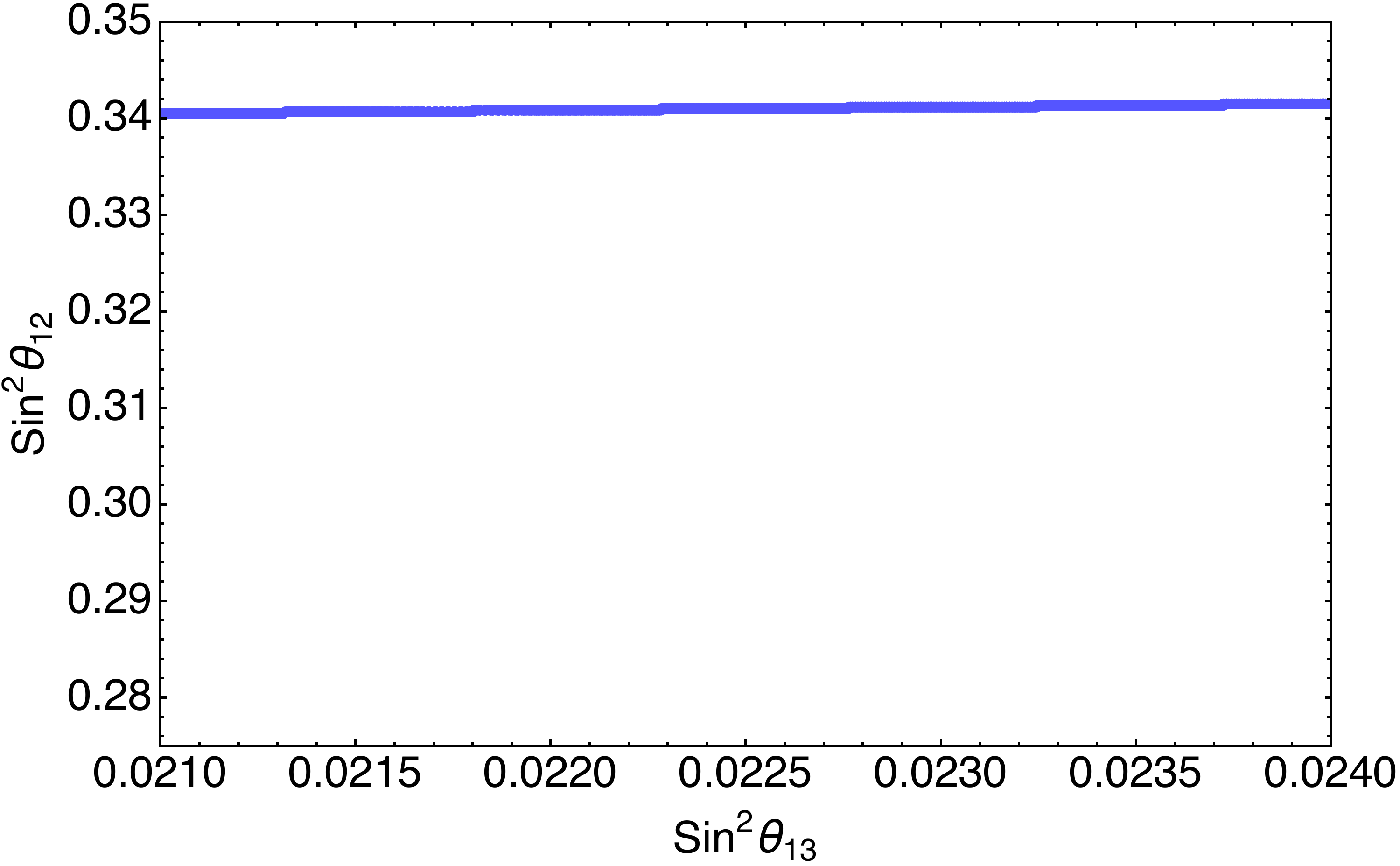}
\includegraphics[scale=0.2]{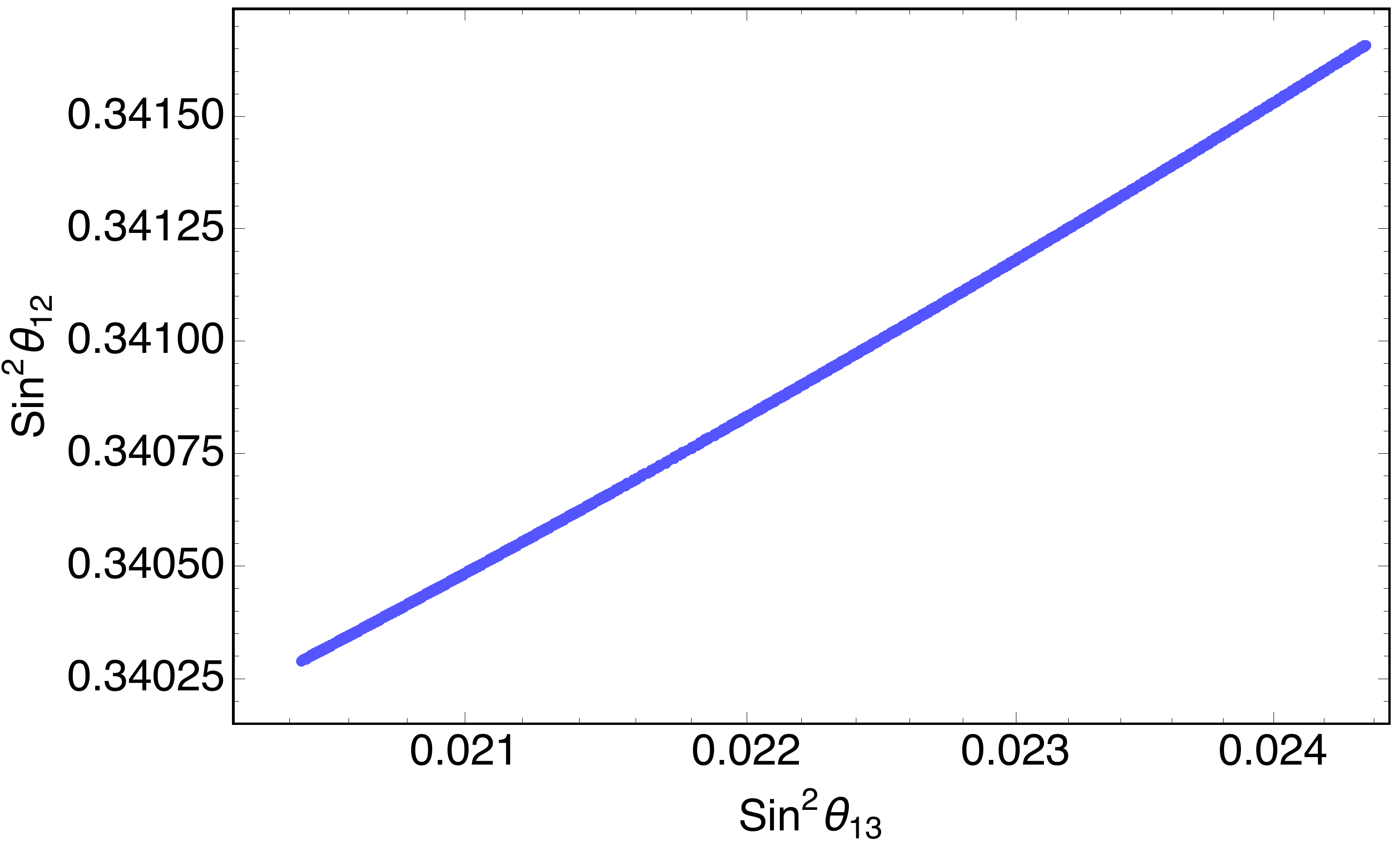}\\
\includegraphics[scale=0.2]{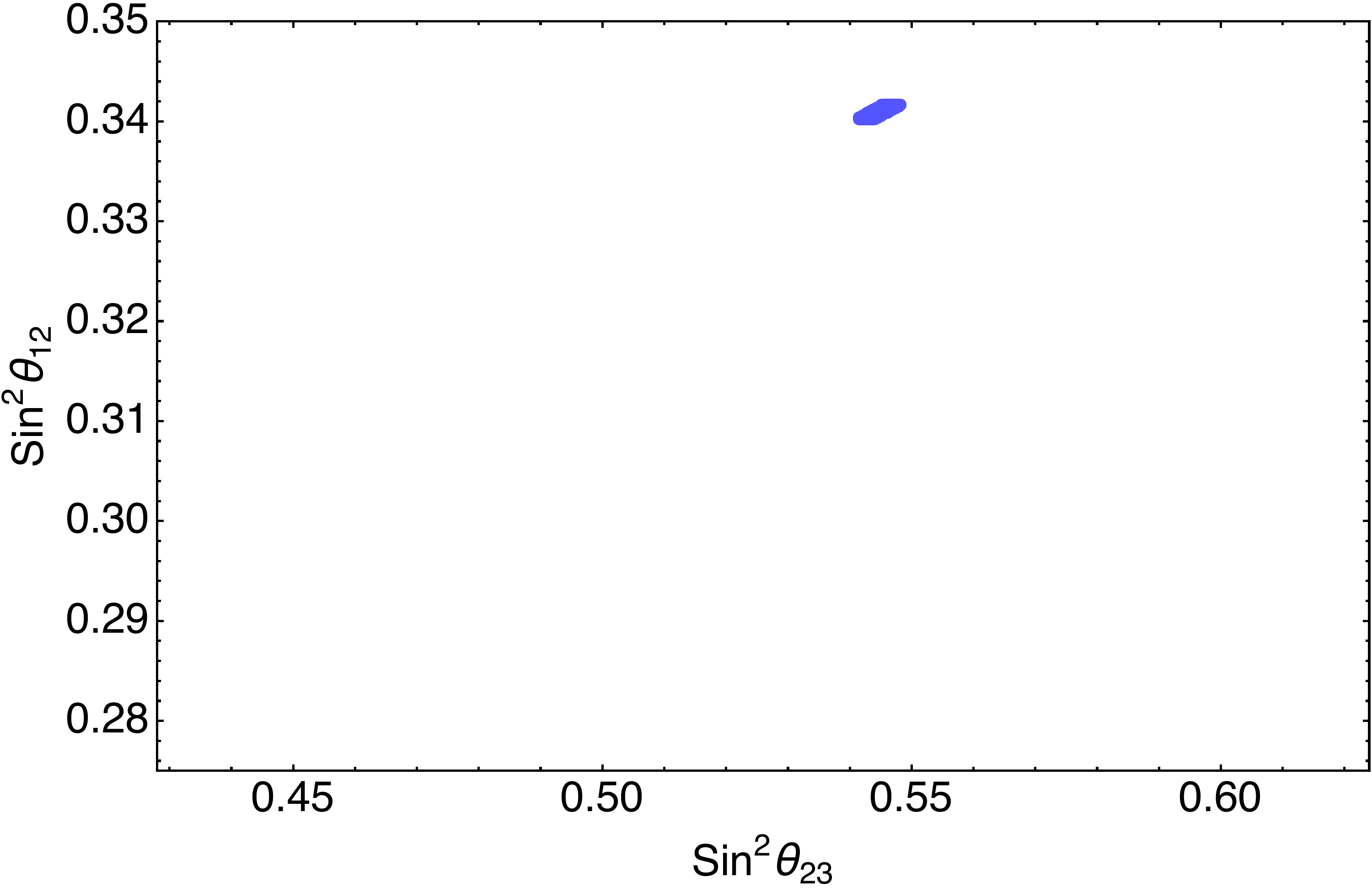}
\includegraphics[scale=0.2]{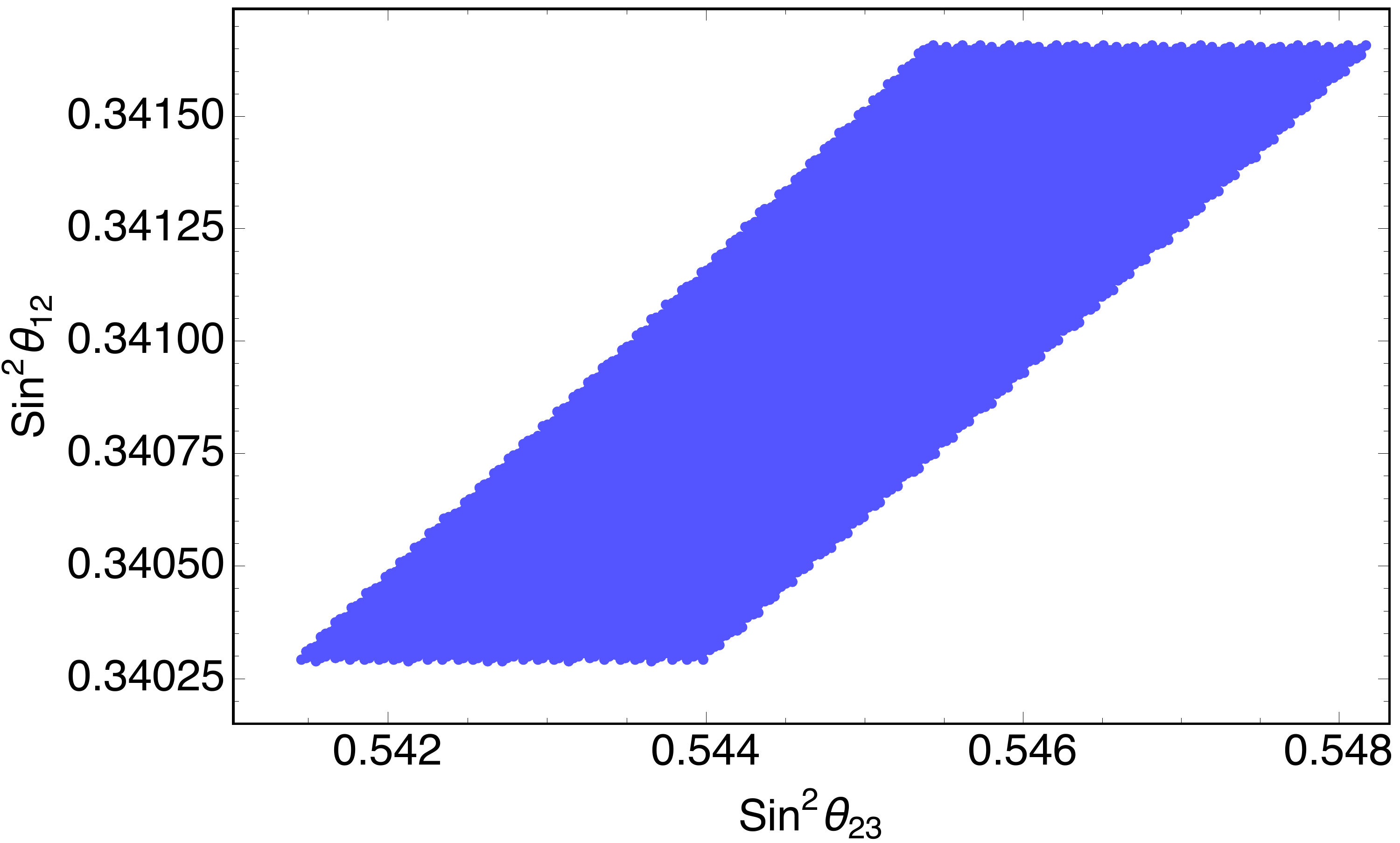}\\
\includegraphics[scale=0.2]{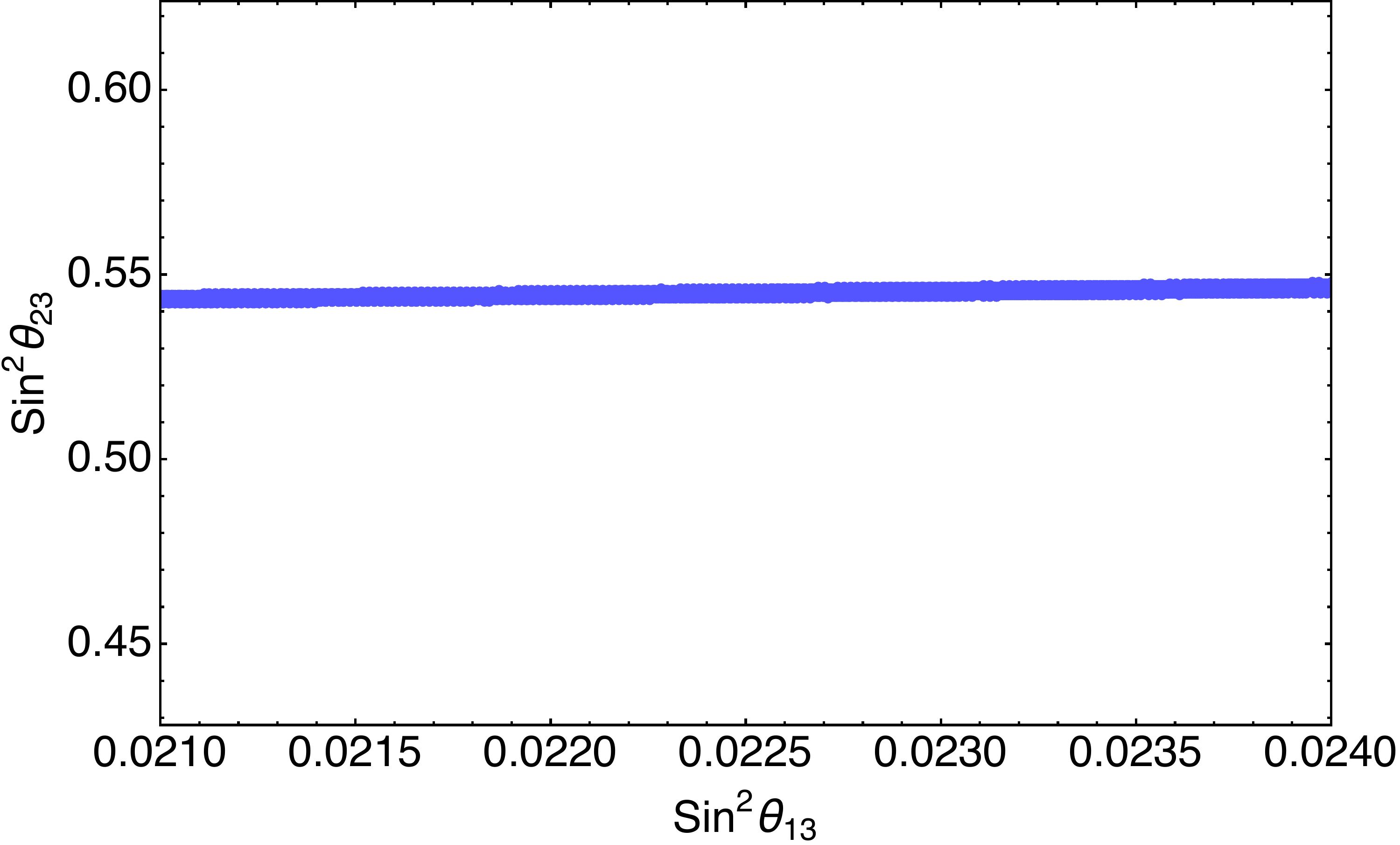}
\includegraphics[scale=0.2]{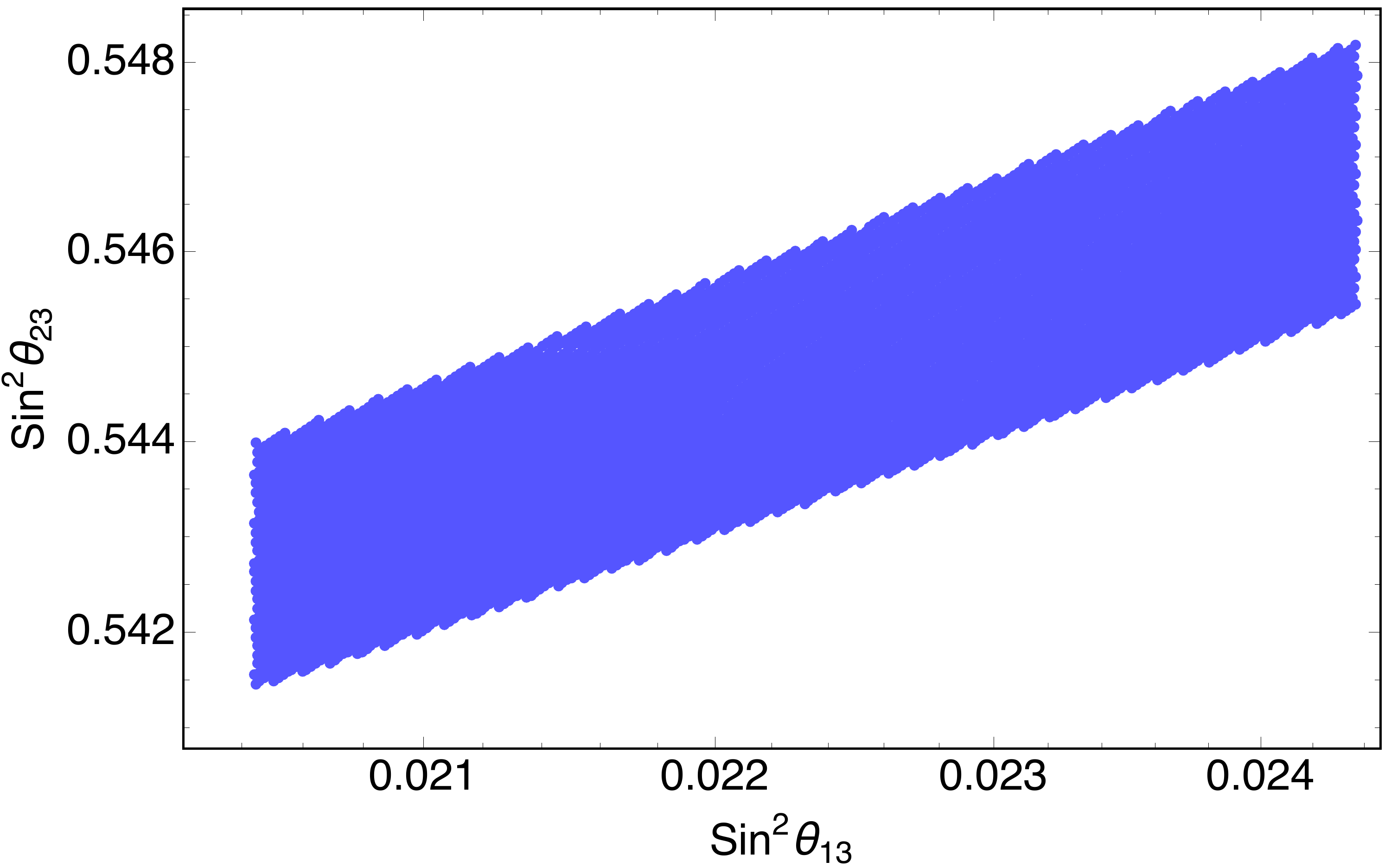}
\caption[]{Correlations among the mixing angles found using equations \ref{eq:t13}, \ref{eq:t12} and \ref{eq:t23}. For an explanation please see the text. In the left panel, the predicted ranges of the observables are plotted against their entire $3\sigma$ ranges. The right panel shows the correlations among the angles which is not clearly visible in the left panel.}
\label{correl}
\end{center}
\end{figure}

\begin{table}[h]
\begin{center}
\begin{tabular}{|c|c|c|}
\hline
& Prediction & Experimental range\\
\hline
$\sin^2 \theta_{12}$ & $ 0.340\rightarrow 0.342$ & $0.275 \rightarrow 0.350$\\
$\sin^2 \theta_{23}$ & $ 0.541\rightarrow 0.548$ & $0.428 \rightarrow 0.624$\\
$\sin \delta$ & $ -0.916\rightarrow-0.905 $ & $-1 \rightarrow 0.707$\\
 $|U_{e4}|^2$ & $0.021\rightarrow 0.038$& $0.012 \rightarrow 0.047$\\
 $|U_{\mu4}|^2$ & $0.007 \rightarrow 0.013$& $0.005\rightarrow 0.03$\\
 $|U_{\tau4}|^2$ & $0.004 \rightarrow 0.008$& $< 0.16$\\
$m_{\beta\beta}$ & $0.0302~\text{eV} \rightarrow 0.0371~\text{eV}$& $<0.05~\text{eV}$\\
\hline
\end{tabular}
\caption{The values of the observables predicted by the model in comparison to their experimental ranges \cite{article, Gariazzo:2015rra, KamLAND-Zen:2016pfg}. $m_{\beta\beta}<0.05~\text{eV}$ is the most stringent bound from the KamLAND-Zen experiment~\cite{KamLAND-Zen:2016pfg}.}
\label{tab:output}
\end{center}
\end{table}

Substituting this range of values in the expression of the solar mixing angle, Eq.~(\ref{eq:t12}), we predict
\be
0.340 \leq \sin^2 \theta_{12}  \leq 0.342. \label{eq:t12predict}
\ee
$\text{TM}_2$ mixing fixes $|U_{e2}|^2$ to be $\frac{1}{3}$. We also have $|U_{e2}|^2=\sin^2 \theta_{12}\cos^2 \theta_{13}$. Therefore, $\text{TM}_2$ scheme strongly constrains $\theta_{12}$ given the precise experimental determination of $\theta_{13}$. The resulting prediction, Eq.~(\ref{eq:t12predict}), is consistent with the $3\sigma$  experimental range $0.275 \leq \sin^2 \theta_{12}  \leq 0.350$, Table~\ref{tab:output}. However, a more precise determination of the solar mixing angle, for instance from
reactor experiments \cite {Bandyopadhyay:2004cp, Bandyopadhyay:2003du} can test this prediction.

Substituting the allowed range of $\ks$ and $\ka$ in the expressions of the atmospheric mixing angle, the Jarlskog invariant and the Dirac CP phase, Eqs.~(\ref{eq:t23}-\ref{eq:sindelta}), we predict
\begin{align}
0.541 \leq &\sin^2 \theta_{23}  \leq 0.548, \label{eq:t23predict}\\
-0.916 \leq &\sin \delta  \leq -0.905 \quad \text{with} \quad -0.0329 \leq J  \leq {\edit -0.0299}. \label{eq:deltapredict}
\end{align}
These predictions are also consistent with the experimental ranges, Table~\ref{tab:output}. Note that the determination of the octant of $\theta_{23}$ is still an open problem experimentally. If the $\mu$-$\tau$ reflection symmetry \cite{Harrison:2002et,Harrison:2002kp,Grimus:2003yn,Feruglio:2012cw,Harrison:2004he,Rodejohann:2017lre} is broken, we have $\theta_{23}$ either in the first or the second octant. The model predicts it to be in the second octant. 

In Figure~\ref{correl}, we show the correlations among the mixing angles resulting from the model. The top panel of the figure shows the $\text{TM}_2$ constraint between the solar and reactor mixing angles, $\sin^2 \theta_{12} \cos^2 \theta_{13}=\frac{1}{3}$. For small $\theta_{13}$, we obtain a linear relationship, $\sin^2 \theta_{12}\simeq \frac{1}{3}(1+\sin^2 \theta_{13})$. The mild positive correlations of the atmospheric angle with the solar and the reactor angles shown in the middle and the bottom panels respectively are nothing but the result of the constraint among these quantities given in Eq.~(\ref{eq:atmconstraint}). We also note that the deviation from maximal atmospheric mixing shown in these plots ($0.041\text{-}0.048$) is consistent with the deviation obtained in Eq.~(\ref{eq:atmconstraint}).

The global fit \cite{article} of oscillation data gives hints for CP violation. Even though the measurement is not precise, $ 135^\circ\leq \delta\leq366^\circ$, it favours a relatively large negative value for $\sin \delta$. Our prediction, Eq.~(\ref{eq:deltapredict}), supports this scenario. In Figure~\ref{fig:t23delta}, we have shown the predictions for $\sin^2 \theta_{23}$ and $\sin \delta$. 

\begin{figure}[tbp]
\begin{center}
\includegraphics[scale=0.75]{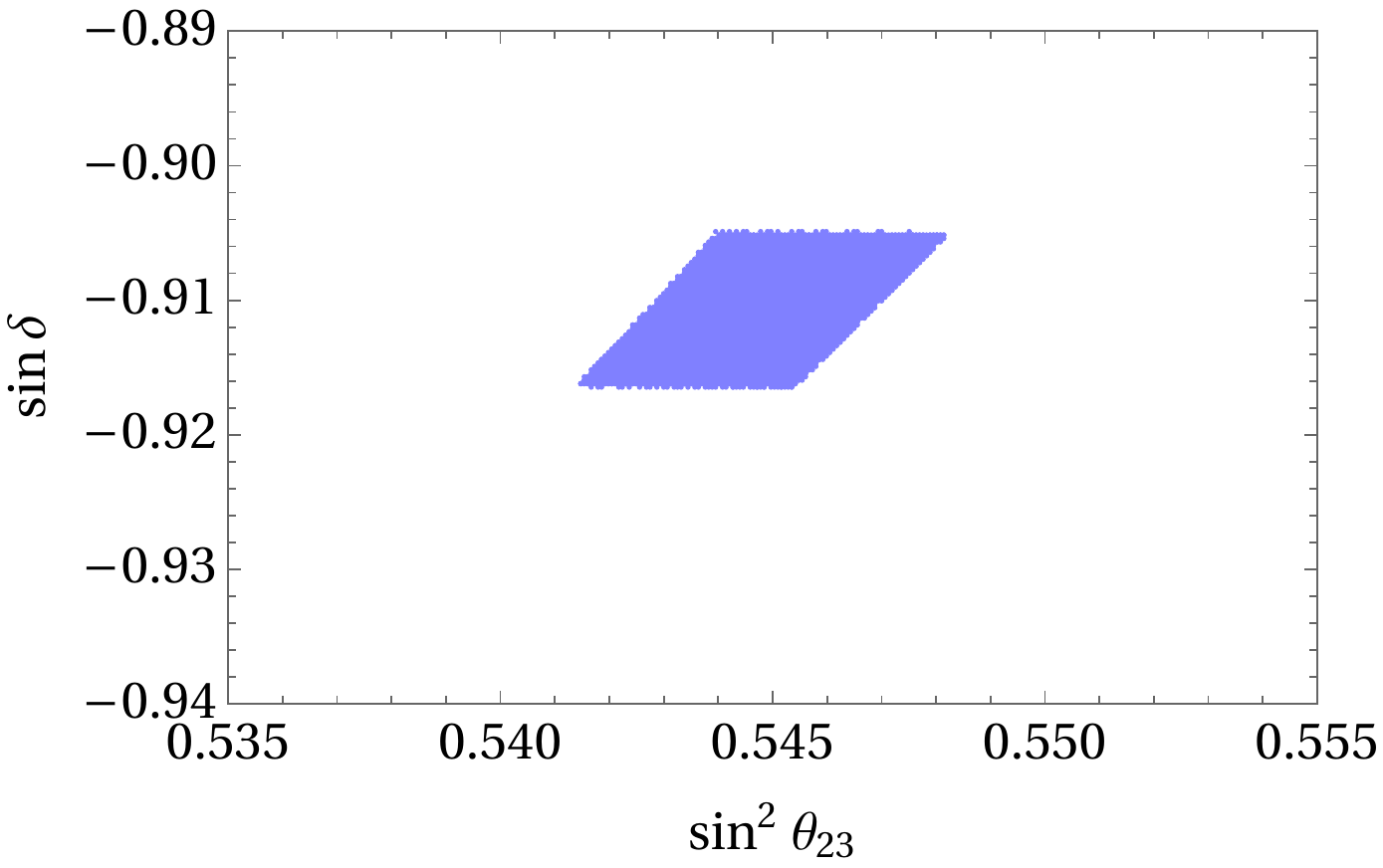}
\caption[]{The predicted ranges of $\sin^2 \theta_{23}$ and $\sin \delta$ as constrained by the parameters $\ks$ and $\ka$.}
\label{fig:t23delta}
\end{center}
\end{figure}

Under the MES scheme, the mass of the lightest neutrino, $m_1$, vanishes\footnote{The higher-order corrections in the MES framework will generate non-zero mass for $m_1$, albeit tiny. Here we ignore this mass. In Appendix~B, we estimate it to be of the order of $10^{-5}$~eV.}. Therefore, the experimental ranges of $\Delta m^2_{21}$ and $\Delta m^2_{31}$, Table~\ref{tab:input}, leads to the prediction,
\be
8.24\times10^{-3}~\text{eV} \leq m_2  \leq 8.95\times10^{-3}~\text{eV}, \quad 4.93\times10^{-2}~\text{eV} \leq m_3  \leq 5.12\times10^{-2}~\text{eV}.
\ee
The model parameter $m$ corresponds to the neutrino mass $m_2$, so its  allowed range is the same as that of $m_2$ given above. Using the range of $\Delta m^2_{41}$ from Table~\ref{tab:input}, we obtain,
\be
0.93~\text{eV} \leq m_4 \leq 1.42~\text{eV},
\ee
which also corresponds to the range of the model parameter, $m_s$, Eq.~(\ref{eq:m4ms}).

The active-sterile mixing observables, Eqs.~\ref{eq:ue4}-\ref{eq:utau4}, depend on all the four model parameters, $\ks$, $\ka$, $m$, and $m_s$. By varying these parameters within their respective ranges we predict the values of these observables, i.e.~$|U_{e4}|^2$, $|U_{\mu4}|^2$ and $|U_{\tau4}|^2$, Table~\ref{tab:output}. These predictions are well within their corresponding experimental ranges. In Figure~\ref{fig:emutau4}, we have plotted them against the parameter, $m_s$. 

\begin{figure}[tbp]
\begin{tabular}{ccc}
  \includegraphics[width=50mm]{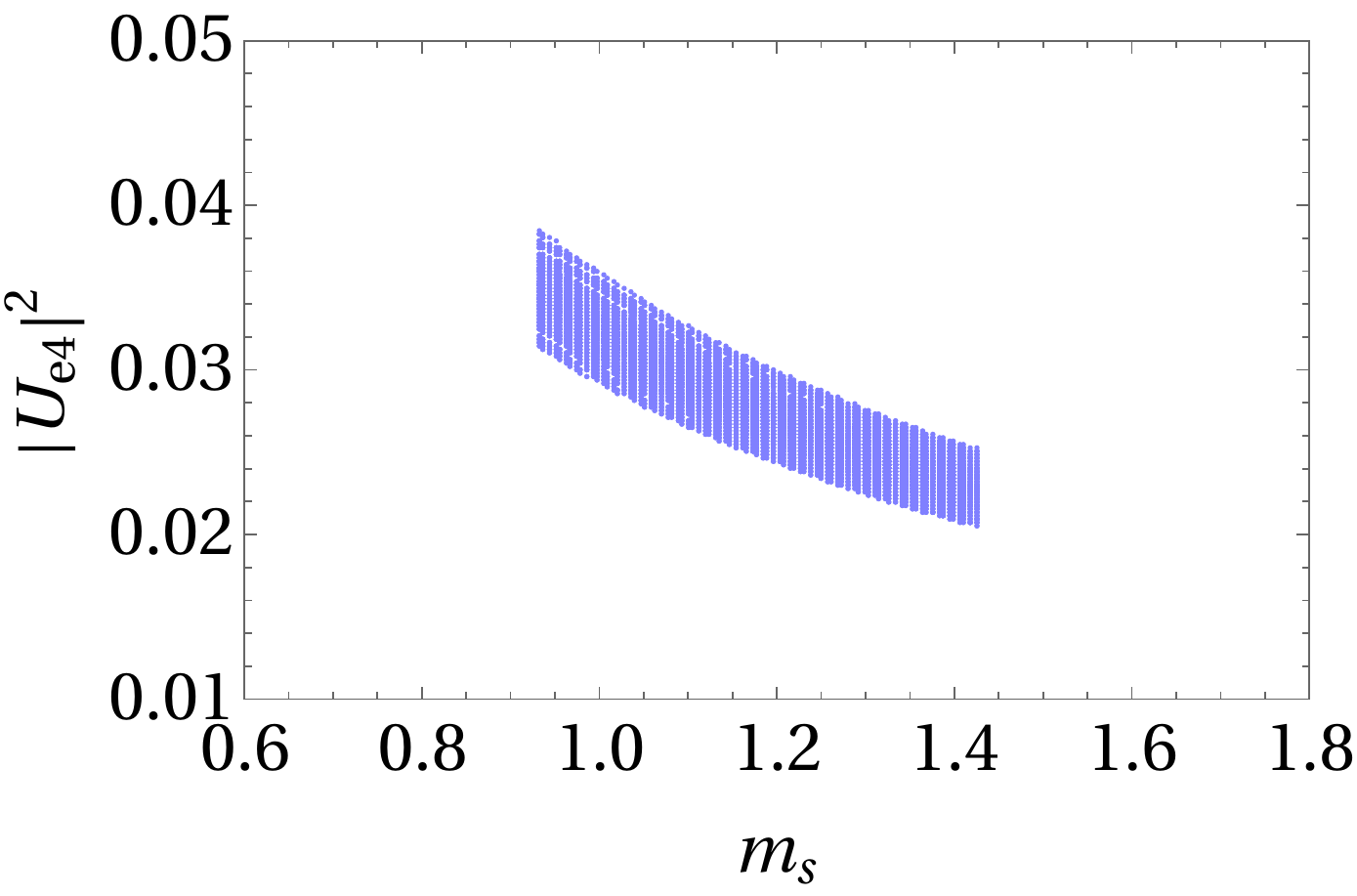} & \includegraphics[width=50mm]{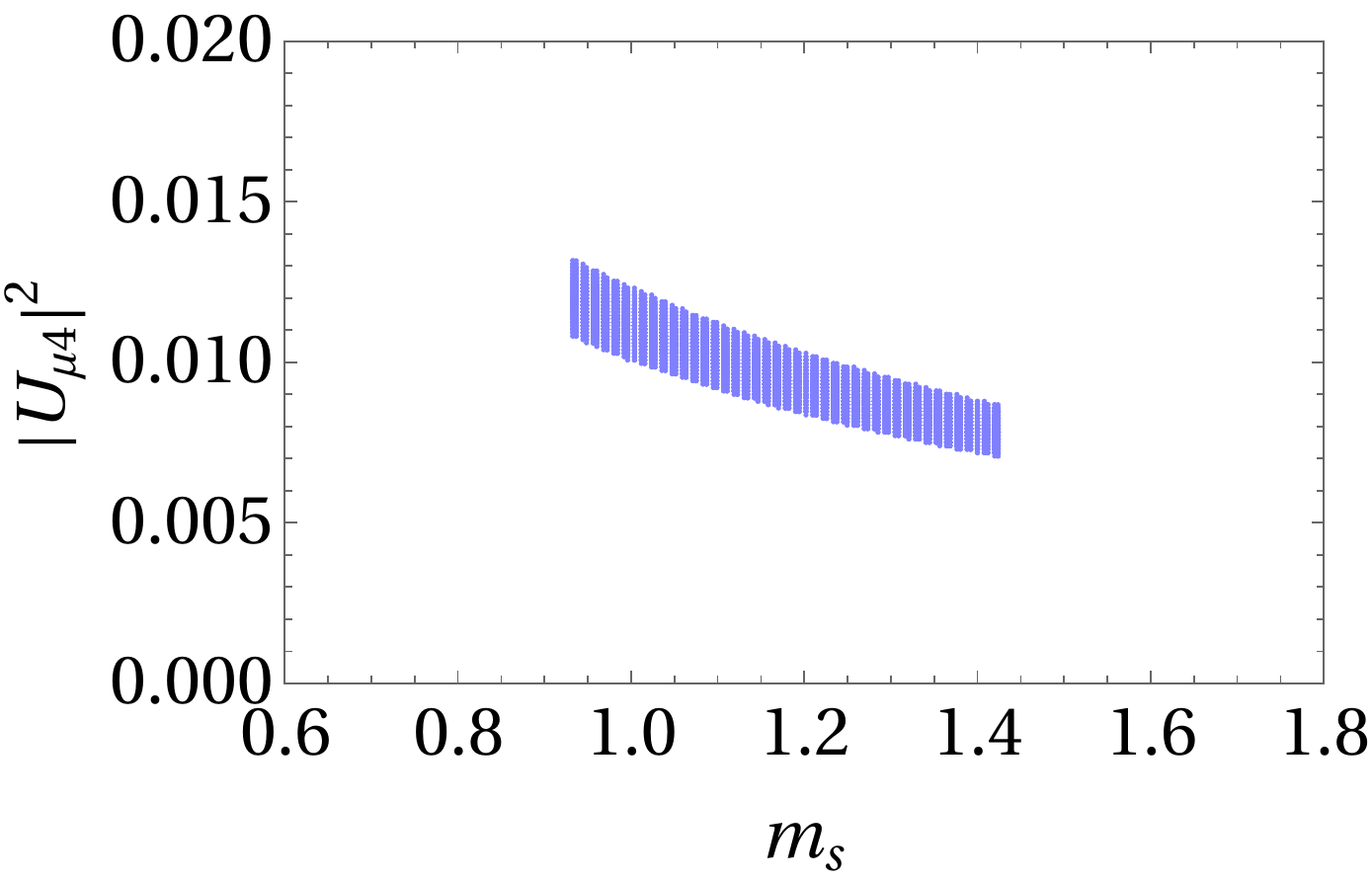}&   \includegraphics[width=50mm]{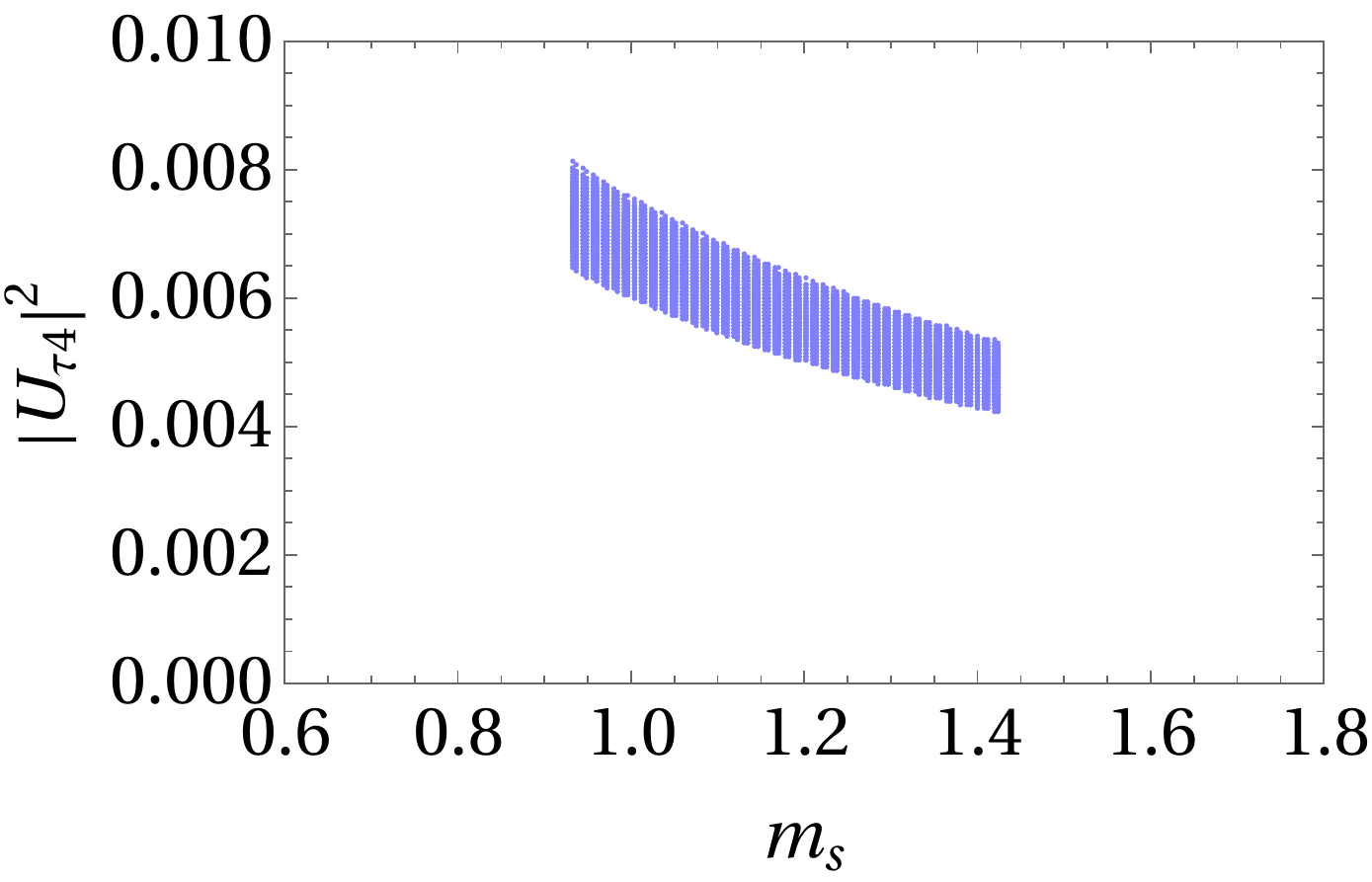} \\
(a)  $|U_{e4}|^2$ vs $m_s$ & (b) $|U_{\mu4}|^2$ vs $m_s$ &(c) $|U_{\tau4}|^2$ vs $m_s$ \\[6pt]
\end{tabular}
\caption{The active-sterile mixing observables predicted by the model plotted against $m_s$. These curves show the inverse relationship between the moduli-squared values of the active-sterile mixing elements and the sterile neutrino mass as can be inferred from Eqs.~(\ref{eq:ue4}-\ref{eq:utau4}).}
\label{fig:emutau4}
\end{figure}

\begin{figure}[tbp]
\begin{center}
\includegraphics[scale=0.4]{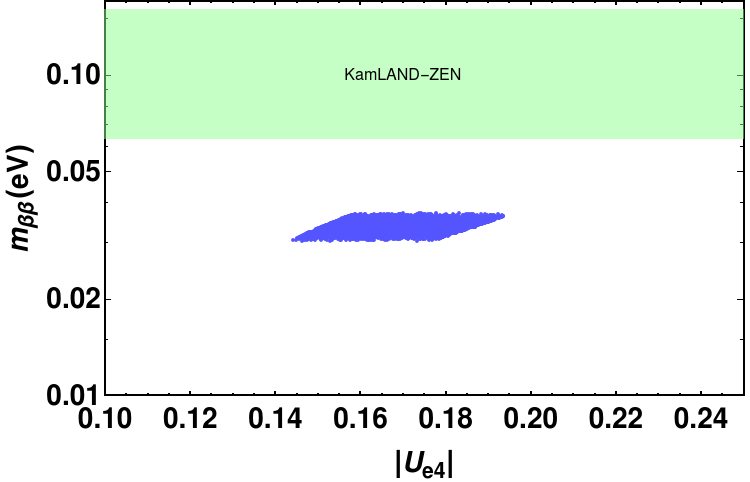}
\caption[]{The prediction of the effective neutrino mass, $m_{\beta\beta}$, in relation to the active-sterile mixing strength.}
\label{fig:ndbd}
\end{center}
\end{figure}

Substituting the allowed ranges of $\ks$, $\ka$, $m$, and $m_s$ in Eq.~(\ref{eq:mbb}), we predict the value of the neutrinoless double-beta decay mass, Table~\ref{tab:output}, which is also shown in Fig.~\ref{fig:ndbd}. This range is quite narrow because the model strongly constrains the first row of the mixing matrix through the parameters $\ks$ and $\ka$, which effectively constrains the Majorana phases as well.

Cosmological observations set upper bounds to the sum of the neutrino masses for three generations of neutrinos, $\Sigma m_i = m_1+m_2+m_3$. However, in the presence of the sterile neutrino, the bound gets affected. At the same time, some recent cosmological models offer an explanation in favour of the existence of the sterile neutrino via the so-called secret interactions. The broken $U(1)_s$ gauge symmetry in our model will lead to a massive gauge boson which can mediate such an interaction. For a detailed discussion about the cosmological implications of sterile neutrinos having secret interactions, please see the references  \cite{Dasgupta:2013zpn,Chu:2018gxk,Mazumdar:2019tbm}. 

\section{Discussion and Conclusion}\label{Conclusion}

In this paper, we construct the leptonic mass matrices in terms of the VEVs of a set of flavon fields transforming under the discrete symmetry group $A_4\times C_4 \times C_6 \times C_2$ and the VEVs of the SM Higgs and a sterile sector Higgs. In the charged-lepton sector, we obtain a non-diagonal mass matrix. In the neutrino sector, we use the MES formula, Eq.~(\ref{eq:M3}), to construct the effective $3\times3$ seesaw mass matrix. The unitary matrices $U_L$ and $U_\nu$ diagonalise the charged-lepton and the neutrino mass matrices respectively. Their product determines the mixing in the active sector, i.e.~$U_\text{PMNS}\simeq U_L U_\nu$, Eq.~(\ref{eq:u33}). In our model, the unitary contribution from the charged-lepton sector ($U_L$) has a $3\times 3$ trimaximal form, Eq.~(\ref{eq:uomega}). On the other hand, the contribution from the active neutrino sector ($U_\nu$) has the form which corresponds to the second flavour eigenstate being equal to the second mass eigenstate as evident from the off-diagonal zeros in $U_\nu$, Eq.~(\ref{eq:unu}). Consequently, the second column of $U_L$ is preserved in the product $U_LU_\nu$ and as a result, we obtain the $\text{TM}_2$ mixing.

$U_\nu$ obtained in the model contains two parameters $\ks$ and $\ka$. These parameters correspond to the symmetric and the antisymmetric parts of the neutrino mass matrix, $M_D$,  which in turn originate from the symmetric and the antisymmetric parts of the tensor product of triplets of $A_4$. If $\ka$ vanishes, $U_\nu$ becomes bimaximal, i.e.
\be
\ka\rightarrow0 \implies U_\nu \rightarrow  \left(\begin{array}{ccc}
       \frac{1}{\sqrt{2}} & 0 &  \frac{-1}{\sqrt{2}}\\
       0 & 1  & 0 \\ 
       \frac{1}{\sqrt{2}} & 0 & \frac{1}{\sqrt{2}}
       \end{array}\right),
\ee
which will lead to tribimaximal (TBM) mixing. The observation of non-zero reactor angle has ruled out TBM. Hence, the parameter $\ka$ plays the vital role of generating the non-zero reactor angle in the model. This role has been emphasized in Ref.~\cite{Borah:2017qdu,}.

We obtain CP violation even though all the free parameters in the model are real. The charged-lepton mass matrix, $M_l$, Eq.~(\ref{CL}), and the neutrino mass matrix, $M_D$, Eq.~(\ref{eq:dirac}), turn out to be complex on account of the complex VEVs $\langle \phil \rangle$ and $\langle \phi \rangle$ respectively. Hence CP is broken spontaneously in the model. Since $M_l$ and $M_D$ are complex, the corresponding diagonalising matrices $U_L$ and $U_\nu$ also become complex and they generate the complex mixing matrix, $U_\text{PMNS}\simeq U_L U_\nu$. It can be shown that if $U_\nu$ were real, the resulting mixing matrix $U_L U_\nu$ would be symmetric under $\mu$-$\tau$ reflection implying $\theta_{23}=\frac{\pi}{4}$. In such a scenario, despite $U_\nu$ being real, CP would be maximally broken ($\delta = \pm \frac{\pi}{2}$) because of the complex contribution from the charged-lepton sector ($U_L$) alone. Our model, with $U_\nu$ also being complex, breaks $\mu$-$\tau$ reflection symmetry and we obtain $\theta_{23}\neq\frac{\pi}{4}$. The complex $U_\nu$ also shifts $\delta$ away from its maximal value, i.e.~$\delta\neq\pm \frac{\pi}{2}$. Therefore, the origin of the non-maximal values of the atmospheric mixing as well as the CP phase is the complex VEV, $\langle \phi \rangle$. 

LSND and MiniBooNE observations suggest the existence of sterile neutrinos. The observed active-sterile mixing ($|U_{e4}|^2$, $|U_{\mu4}|^2$) is found to be of the order of $\frac{\sqrt{\Delta m^2_{21}}}{\sqrt{\Delta m^2_{41}}}$. The Minimal Extended Seesaw provides a natural framework to achieve this relationship. It is in this context that we built the model to explain both the active and the sterile mixing observables. In the model, these observables are given in terms of four parameters, $\ks$, $\ka$, $m$ and $m_s$. We use the experimental ranges of the reactor mixing angle, $\sin^2\theta_{13}$, and the mass-squared differences, $\Delta m^2_{21}$ and $\Delta m^2_{31}$, to extract the allowed values of $\ks$, $\ka$ and $m$, as we obtain $m_2 < m_3$ corresponding to normal hierarchy. The extracted values of $\ks$ and $\ka$ are used to predict $\theta_{23}$ and $\delta$. These predictions can be tested when these observables are measured more precisely in future oscillation experiments. The model parameter, $m_s$, corresponds to the sterile neutrino mass and is determined by the active-sterile mass-squared difference, $\Delta m^2_{41}$. The three model parameters, $\ks$, $\ka$ and $m$ (constrained using $\sin^2\theta_{13}$, $\Delta m^2_{21}$ and $\Delta m^2_{31}$), as well as the fourth parameter, $m_s$ (constrained using $\Delta m^2_{41}$), are used to evaluate the active-sterile mixing. We find that these values are consistent with the experimental results. We also obtain strong constraints on the range of the effective mass governing the neutrinoless double-beta decay. 
  
%---------Appendix-----------------

\appendix
\section{Uniquely defining the flavon VEVs}
\label{sec:appendixa}

The model contains the flavon multiplets, $\phil$, $\eta$, $\phi$, $\phis$ and $\eta_\nu$. The alignments of their VEVs in the flavour space play a crucial role in determining the model's phenomenology. In this Appendix, we provide justifications for these alignments by uniquely defining them using symmetries.

The flavon $\phil$ couples in the charged-lepton sector. Consider the alignment $\langle \phil \rangle \propto (1,\ob,\om)^T$, Eq.~(\ref{eq:vevphil}). This VEV is invariant under the following group action:
\be\label{eq:clres}
\om T \langle \phil \rangle = \langle \phil \rangle,
\ee
where $\om$ and $T$ are elements of $C_3$ and $A_4$ respectively under which $\phil$ transforms. In other words, $\om T$ generates the residual symmetry of $\langle \phil \rangle$ and this symmetry uniquely defines $\langle \phil \rangle$ (up to multiplication with a constant\footnote{This constant, i.e.~$v_l$ in Eq.~(\ref{eq:vevphil}), can be complex in general. Note that $v_l$ being complex does not alter the phenomenology of the charged-lepton sector.}). A triplet flavon whose VEV is defined by the residual symmetry, Eq.~(\ref{eq:clres}), was recently utilised in the construction of the charged-lepton mass matrix in Ref.~\cite{Krishnan:2019xmk}.

The singlet $\eta$ and the triplet $\phi$ couple in the neutrino Dirac sector. The VEV of the singlet, Eq.~(\ref{eq:veveta}),  is assumed to be real. This can be ensured by assuming the residual symmetry under conjugation,
\be
\langle \eta\rangle ^* =\langle \eta\rangle.
\ee
The VEV $\langle \phi\rangle \propto (0,-i,0)^T$, Eq.~(\ref{eq:vevphi}), is uniquely defined using the residual symmetries,
\be
T^2 S T \langle \phi \rangle = \langle \phi \rangle, \quad  -1 \langle \phi \rangle^* = \langle \phi \rangle.
\ee  
These symmetries ensure that the first and the third components of $\langle \phi \rangle$ vanish and the phase of the second component is $-i$. Note that the constant of proportionality in the VEV, i.e.~$v_\phi$ in Eq.~(\ref{eq:vevphi}), is real.

The triplet $\phis$ couples in the sterile sector. Its VEV, $\langle \phis \rangle  \propto (1,0,1)^T$, Eq.~(\ref{eq:vevphis}), has the following residual symmetries:
\be
-1 \,\, T^2 S T \langle \phis \rangle = \langle \phis \rangle, \quad  \langle \phis \rangle^* = \langle \phis \rangle.
\ee  
The VEV of the $A_4$ triplet in the form $\propto (1,0,1)^T$ has been widely used in the literature \cite{King:2006np,Felipe:2013vwa,Pramanick:2017wry}. This alignment can be uniquely obtained from the symmetric tensor product involving the alignments $(0,1,0)^T$ and $(1,1,1)^T$\footnote{The alignment $(1,1,1)^T$ can be uniquely defined by the residual symmetry, $T (1,1,1)^T=(1,1,1)^T$.},
\be
\left((0,1,0)^T,(1,1,1)^T\right)_{\rt s} \propto (1,0,1)^T.
\ee

The VEV of the singlet $\eta_\nu$ can be trivially assigned any complex constant since its phase has no effect on the model's phenomenology.

Here we have assumed that various flavon VEVs have distinct residual symmetries. It is interesting to note that if all the VEVs are taken together, none of these symmetries survive. A comprehensive study of the origin of the VEVs requires the construction of the flavon potential. If different irreducible multiplets are decoupled in the potential, then it can lead to the corresponding VEVs having separate residual symmetries. The construction of a potential with an inbuilt mechanism to ensure the decoupling of the irreducible multiplets is beyond the scope of this work. However, we assume that our VEVs are generated from such a potential.

\section{Numerical verification of approximations}
\label{sec:appendixb}

In this Appendix, we construct the charged-lepton and the neutrino mass matrices using a representative set of model parameters and numerically extract the masses and the mixing observables without employing approximations. Thus we verify the correctness of the various results obtained in the main body of the paper.  In our calculations, we use
\be\label{eq:flavvevs}
\Lambda = 10^{13}, \quad v_x = 10^{10}, 
\ee
where $v_x$ represents $v_l, v_\eta, v_\phi, v_s$ and $v_\nu$. All mass units are given in GeV, unless otherwise specified. The SM Higgs and the sterile Higgs are given the following VEVs,
\begin{align}
v&=176, \label{eq:higgsvev1}\\
v'&= 2000\label{eq:higgsvev2}.
\end{align}
Substituting Eqs.~(\ref{eq:flavvevs}, \ref{eq:higgsvev1}) in Eq.~(\ref{CL}), we obtain the charged-lepton mass matrix in the form,
\be\label{CLsub}
       M_l = 176\times 10^{-6}\left(\begin{array}{ccc}
      (1.7-i {\edit 0.9})Y_e & 0 & 0 \\
       (0.8-i {\edit 1.5})Y_e & 0 & 0  \\ 
       (1.7-i{\edit 0.9})Y_e & 0  & 0
       \end{array}\right)+176\times 10^{-3}\left(\begin{array}{ccc}
       0 & Y_\mu & Y_\tau \\
       0 & \om Y_\mu & \ob Y_\tau  \\ 
       0 & \ob Y_\mu  & \om Y_\tau
       \end{array}\right).          
\ee
Since several higher-order flavon triplets contribute in the construction of the electron mass term, there will be a corresponding set of Yukawa-like free parameters. We assign random real numbers to these parameters resulting in the first column of $M_l$, Eq.~(\ref{CLsub})\footnote{The exact values of these numbers are irrelevant in an order-of-magnitude calculation.}. For convenience, we have introduced a parameter $Y_e$ in this column. Diagonalising $M_l$ using the unitary matrices $U_L$ and $U_R$, Eq.~(\ref{eq:cldiagonalisation}), produces the charged-lepton masses.  Their experimental values ($m_\tau=1777$~MeV, $m_\mu=106$~MeV and $m_e=0.511$~MeV) are obtained with the substitution,
\be
Y_\tau= \frac{1}{\sqrt{3}} \frac{1}{176} 1777=5.83, \quad Y_\mu= \frac{1}{\sqrt{3}} \frac{1}{176} 106 = 0.348, \quad Y_e= 0.942.
\ee
These coupling constants are of the order of one as we expect. The corresponding diagonalising matrices are 
\begin{align}
  U_L &= \left(\begin{array}{ccc}
       0.577 & 0.577 & 0.577 \\
       0.577 & -0.289-i 0.500 & -0.289+i 0.500  \\ 
       0.577 & -0.289+i 0.500  & -0.289-i 0.500
       \end{array}\right),\\
 U_R&=\left(\begin{array}{ccc}
       {\edit 0.786-i 0.618} & {\edit 0.001 + i 0.001} & 0.000 \\
       {\edit i 0.001} & 1.000 & 0.000  \\ 
       0.000 & 0.000  & 1.000
       \end{array}\right).          
\end{align}
We find that the above calculation of $U_L$ is consistent with the expression given in Eq.~(\ref{eq:uomega}) within a deviation of the order of $\frac{v_x^2}{\Lambda^2}$.

To construct the neutrino mass matrices, we make the following assignments:
\be\label{eq:allyukawas}
Y_\eta=1.67, \quad Y_{\phi s}=3.59, \quad Y_{\phi a}=0.80, \quad Y_s =1.23, \quad Y_\nu = 1.00.
\ee
Substituting Eqs.~(\ref{eq:flavvevs}-\ref{eq:higgsvev2}, \ref{eq:allyukawas}) in Eqs.~(\ref{eq:dirac}-\ref{eq:majorana}), we obtain 
\be\label{eq:subm1}
M_D = \left(\begin{array}{ccc}
       0.294 & 0 & -i0.491 \\
       0 & 0.294 & 0\\
       -i0.773 & 0 & 0.294
       \end{array}\right), \quad M_s = {\edit(2.46, 0, 2.46)}, \quad M_R = \text{diag}(1,1,1)10^{10}.
\ee
We also introduce mass matrices involving $\bar \nu_L \nu_L^c$, $\bar \nu_L \nu_s$ and $\bar \nu_s^c \nu_s$, which are highly suppressed as described at the end of Section~3,  
\be\label{eq:subm3}
\bar \nu_L \nu_L^c\text{:} \quad M_1 = \left(\begin{array}{ccc}
       1.24 & -2.32-i4.02 & -1.39+i4.02 \\
       -2.32-i4.02 & 1.24 & 4.65\\
       -1.39+i4.02 & 4.65 & 1.24
       \end{array}\right)10^{-15},
\ee
\be\label{eq:subm4}
\bar \nu_L \nu_s\text{:} \quad M_2 = \left(\begin{array}{c}
       2.11-i1.41  \\
       0 \\
       2.11+i2.82
       \end{array}\right)10^{-14}, \quad \bar \nu_s^c \nu_s\text{:} \quad M_3= \left(\begin{array}{c}
       2.80
       \end{array}\right)10^{-13}.
\ee
Here also, the matrices have been populated with a random choice of Yukawa-like couplings corresponding to the higher-order terms. In terms of Eqs.~(\ref{eq:subm1}-\ref{eq:subm4}), we construct the neutrino mass matrix in the `$\nu=(\nu_L, \nu^c_s, \nu^c_R)$'-basis:
\be
\bar \nu \nu^c\text{:} \quad M_\nu^{7 \times 7} = \left(\begin{array}{ccc}
       M_1 & M_2 & M_D \\
       M_2^T & M_3 & {\edit M_s}\\
       M_D^T & {\edit M_s^T} & M_R
       \end{array}\right). 
\ee
This matrix is diagonalised to obtain the neutrino masses,
\begin{align}
\begin{split}
U_\nu^{7 \times 7 ^\dagger} M_\nu^{7 \times 7} \,\,U_\nu^{7 \times 7^*} =&\\ &\hspace{-3cm}\text{diag}\left(1.4\times 10^\text{-14},8.6\times 10^\text{-12},5.0\times 10^\text{-11},1.3\times 10^\text{-9},1.0\times 10^\text{10},1.0\times 10^\text{10},1.0\times 10^\text{10}\right),
\end{split}
\end{align}
where
\be
U_\nu^{7 \times 7} = \left(\begin{array}{cc}
       U_\nu^{4 \times 4} & \mathcal{O}(10^{-10}) \\
       \mathcal{O}(10^{-10}) & U_\nu^\text{heavy}
       \end{array}\right).
\ee
The structure of $U_\nu^\text{heavy}$ depends on the higher-order corrections to $M_R$ which break the degeneracy of the heavy neutrino masses. These corrections are of no significance to the present work. We note that the mass acquired by the lightest neutrino is of the order of $M_2=\mathcal{O}(10^{-14})$. 

The $4\times4$ light-neutrino mixing matrix is obtained as
\be
U=   \left(\begin{array}{cc}
       U_L & 0 \\
       0 & 1
       \end{array}\right) U_\nu^{4 \times 4}= \left(\begin{array}{cccc}
       -0.332+i0.713 & 0.571+i0.086& -0.059+i0.137& 0.067-i0.146 \\
       -0.323+i0.154 & -0.212-i0.537& -0.438-i0.581& 0.096+i0.017\\
	0.157+i0.441&-0.359+i0.452&0.373-i0.551& -0.060-{\edit i}0.039\\
	0.185& 0.000&0.079&0.980
       \end{array}\right),
\ee
This matrix is consistent with the results we obtained using the seesaw approximations.

\bibliographystyle{JHEP}
\bibliography{mesrefs}

\end{document}